\documentclass[onecolumn,usenatbib,amsmath,amssymb,amsbsy]{mn2e}

\usepackage{amsbsy}
\usepackage{amssymb}
\usepackage{graphicx}
\usepackage{natbib}
\def\bv{\bar{v}}
\def\hom{\hat{\Omega}}

\def\X{{\mathrm{x}}}
\def\Y{{\mathrm{y}}}

\def\n{{\rm n}}
\def\p{{\rm p}}

\def\c{{\rm c}}

\def\be{\begin{equation}}
\def\ee{\end{equation}}
\def\bea{\begin{eqnarray}}
\def\eea{\end{eqnarray}}

\def\bear{\begin{eqnarray}}
\def\eear{\end{eqnarray}}
\def\non{\nonumber \\}

\newcommand{\Mat}[4]{\left( \begin{array}{cc} {#1} & {#2} \\ {#3} & {#4} \end{array} \right)}
\newcommand{\pd}[2]{\frac{\partial {#1}}{\partial {#2}}}
\newcommand{\vc}[1]{\boldsymbol{#1}}

\newcommand{\Vec}[2]{\left( \begin{array}{c} {#1} \\ {#2} \end{array} \right)}

\begin{document}

\title{Waves and instabilities in dissipative rotating superfluid neutron stars}

\author[T. Sidery et al]{ T. Sidery$^1$, N.~Andersson$^1$ and G.L. Comer$^2$ \\
$^1$ School of Mathematics, University of Southampton,
Southampton SO17 1BJ, United Kingdom\\
$^2$ Department of Physics \& Center for Fluids at All Scales,
Saint Louis University, St.~Louis,
MO, 63156-0907, USA}

\maketitle

\begin{abstract}
We discuss wave propagation in rotating superfluid neutron star cores, 
taking into account the vortex mediated mutual friction force. For models where the two fluids co-rotate in the 
unperturbed state, our analysis clarifies the role 
of chemical coupling and entrainment for sound and inertial waves. We
also investigate the mutual friction damping, providing results that demonstrate the well-known fact that 
sound waves propagating along a vortex array are undamped. We  show that the same is not true 
for  inertial waves, which are damped by the mutual friction regardless of the propagation direction. 
We then include the vortex tension, which arises due to local vortex curvature.
Focussing on purely transverse inertial waves, we derive the small correction that the tension induces in the wave 
frequency. Finally, we allow for a relative linear flow in the background (along the rotation axis).   
In this case we show how the mutual friction coupling may induce a dynamical instability in the
inertial waves. We discuss the critical flow required for the instability to be present, its
physical interpretation and the possible relevance it may have for neutron star physics.   
\end{abstract}

\section{Introduction}

Our understanding of neutron star interiors currently relies almost entirely on
observations of the ``surface''. In some cases we have  upper limits on the temperature,
which can be combined with an estimated age to yield information about neutron star cooling.
This in turn depends on the interior physics, e.g. whether the core contains superfluid components or
not \citep{mincool}. Other evidence comes from
the way that the neutron star
interacts with its environment, e.g. how the magnetosphere affects the spin-down of an isolated radio pulsar.
It is, however, much harder to link this information to the properties of the interior.
It is also difficult to draw definite conclusions about the nature of the core fluid from  bulk quantities such as mass and radius.

The currently most potent tests of our theoretical ideas are provided by observed crust oscillations
in the tails of magnetar giant flares \citep{watts}, and glitches in the spin-down of radio pulsars \citep{glitch}.
It seems plausible that the crust motion in the magnetar events will to some extent depend on the core physics, e.g. whether
the magnetic field penetrates the core or not \citep{gsa,levin}. Meanwhile, the glitches remain the strongest indication
that the core contains partially decoupled superfluid components that may (probably following the onset of some instability)
transfer angular momentum to the crust. In order to improve our models of these events we need
to understand the dynamics of large-scale superfluid systems. One key question that must be addressed if we want to
be able to compare our  models to real data concerns how energy is dissipated in the system. Consider the (relatively simple!)
case when the internal fluid is a mixture of superconducting protons, electrons, and superfluid neutrons.
The equations of motion for such  multi-constituent fluids have been formulated \citep{PRIX} and
constrained to a three fluid model \citep{NILSCOMER}.
The three fluids are neutrons, entropy and charged particles.
It has recently been argued that for this system there are 19, more or less unknown,
dissipation coefficients \citep{NILSCOMER}. This problem is clearly much more intricate than the
standard single fluid case, where one need only worry about shear- and bulk viscosity.

A natural way to gain insight into the nature of a fluid system is to carry out a local analysis of
wave propagation. Such a plane-wave study should provide a better understanding of energy
dissipation in the system, and could perhaps also help constrain
the different parameters. Given our current understanding it is natural
to divide this effort into a number of steps.  This first study is focussed on the
two-constituent model which applies in the low-temperature limit. In this limit, we
know from results for superfluid Helium that the main dissipation mechanism is due to the presence of rotational vortices.
The vortices induce a mutual friction between the two constituents.
We have recently discussed the associated force for neutron stars \citep{ASC1}, including the important
effect of entrainment.
The entrainment is due to the multi-fluid aspects of the problem. When the energy of the system is allowed to depend upon the relative velocities of the constituents,
each constituent momentum is modified in such a way that it is no longer parallel with the corresponding transport velocity.
The model has since been extended to include curved vortices and a suggested form for the average force in a polarized state of 
turbulence \citep{ASC2}.

Since we now understand some of the effects that should be present in a
dissipative superfluid neutron star, it is interesting to ask how they affect motions in the fluid. Hence, we will
study local oscillations in the fluid in various situations.  As the problem soon becomes intractable unless one
introduces a number of simplifying assumptions, we aim to simplify the problem to the point where
we can carry out the analysis analytically. Nevertheless, we believe that our results shed new light on
the dynamics of these systems. Through various dispersion relations and  the properties of the
associated waves, we  get a better understanding of the role of entrainment and the nature of two-fluid inertial waves. 
We  also learn how mutual friction damps both acoustic and inertial waves.
Finally, we  account for the vortex tension and demonstrate how the mutual friction may induce an
instability in the inertial waves. This is the first demonstration of the so-called Donnelly-Glaberson
instability \citep{Glaberson} in a neutron star. We show that the instability belongs to the
general class of two-stream instabilities \citep{twostream}, and discuss how it may trigger turbulence in the neutron star core.

\section{The single fluid problem}

Before  considering the superfluid problem we will discuss the simpler case of a one fluid system.
This is a fairly standard analysis, but it provides a number of results that will be of use later.
Most importantly, it allows us to discuss the nature of inertial waves. We also have an opportunity to
establish the notation and the approach to the plane-wave problem that will be used throughout the paper.

\subsection{The plane-wave approach}

The equations of motion for a single fluid in a rotating frame are, in a coordinate basis,
\be
\left( \pd{}{t} + v^{j} \nabla_{j} \right) v_{i} + 2 \epsilon_{ijk} \Omega^{j} v^{k} + \nabla_{i} (\tilde{\mu} + \Phi) = 0 \ .
\ee
The velocity is given by $v_{i}$, $\Omega_{i}$ is the angular velocity of the frame and $\tilde{\mu}$ is the chemical potential of the fluid
per unit mass. The gravitational potential is represented by $\Phi$. Throughout our analysis we will assume that the effects of gravity
can be neglected. In effect, we make the Cowling approximation by ignoring variations in $\Phi$.
The continuity equation is
\be
\pd{\rho}{t} + \nabla_{j} ( \rho v^{j}) = 0 \ , 
\ee
where $\rho$ is the mass density. Assuming that the background motion is represented by solid bulk rotation, the
corresponding velocity field vanishes in the rotating frame.
Perturbing the system we then have
\be
\delta v_{i} = \bar{v}_{i} e^{i (\omega t + k_j x^j)} \ , 
\ee
i.e. the perturbation is written in the  form of a plane wave, with frequency $\omega$ and wave vector $k_{i}$.
Similarly we have for the perturbed density
\be
  \delta \rho = \bar{\rho} e^{i (\omega t + k_j x^j)} \ , 
\ee
The amplitudes  $\bar{v}_{i}$ and $\bar{\rho}$ will be taken to be constant throughout the analysis.
With these assumptions, the continuity equation becomes
\be
\label{eq:PertContSingle}
   \omega \bar{\rho} +  \rho k_{j} \bar{v}^{j} = 0 \ , 
\ee
where $\rho$ now represents the background density.

To perturb the equations of motion we assume a one-parameter equation of state. Representing the equation of state
by an energy functional $E=E(\rho)$ we then have
\be
\tilde{\mu} = \pd{E}{\rho}
\longrightarrow
\delta \tilde{\mu} = \pd{\tilde{\mu}}{\rho} \delta \rho = c_s^2 { 1 \over \rho} \delta \rho \ ,
\ee
where the sound speed is defined as
\be
c_s^2 = \rho \pd{\tilde{\mu}}{\rho} \ . 
\ee
It now follows that
\be
\label{eq:PertEquationSingle}
 i \omega \bar{v}_{i} + 2 \epsilon_{ijk} \Omega^{j} \bar{v}^{k} + i k_{i} c_s^2 {1 \over \rho} \bar{\rho} = 0 \ . 
\ee
Substituting the continuity equation (\ref{eq:PertContSingle}) into (\ref{eq:PertEquationSingle}) we finally find that
\be \label{eq:PurtEqnSingle}
i \omega \bar{v}_{i} + 2 \epsilon_{ijk} \Omega^{j} \bar{v}^{k} - i c_s^{2} k_{i} \frac{ k_{j} \bar{v}^{j}}{\omega} = 0 \ .
\ee

\subsection{Finding the dispersion relation}

The derivation of the required dispersion relation is straightforward. It is preferable to
work with scalar equations. Hence we first contract (\ref{eq:PurtEqnSingle}) with $k^{i}$. Rearranging the result
we have
\be
\label{eq:PertSingle1}
  i (k^{i} \bar{v}_{i}) \left(\omega - \frac{c_s^2 k^2}{\omega} \right)  - 2 \Omega^{i} \epsilon_{ijk} k^{j} \bar{v}^{k} = 0 \ .
\ee
Contracting (\ref{eq:PurtEqnSingle}) with $\Omega_{i} \epsilon^{ijk} k_{j}$ then leads to
\be
\label{eq:PertSingle2}
  i \omega ( \Omega^{i} \epsilon_{ijk} k^{j} \bar{v}^{k}) + 2  \Omega^{2} ( k_{j} \bar{v}^{j} ) - 2 (k_{j} \Omega^{j}) ( \Omega_{i} \bar{v}^{i}) = 0 \ .
\ee
Finally, contracting (\ref{eq:PurtEqnSingle}) with $\Omega^{j}$ gives
\be
\label{eq:PertSingle3}
  \bar{v}_{j} \Omega^{j} = \frac{c_s^{2}}{\omega^2} (k_{j} \Omega^{j}) ( \bar{v}_{l} k^{l} ) \ .
\ee
Defining the angle $\theta$ such that $  k_{j} \Omega^{j} = k \Omega \cos{\theta}$
we can now use (\ref{eq:PertSingle2}) and (\ref{eq:PertSingle3}) to substitute for
$\epsilon_{ijk} \Omega^{i}  k^{j} \bar{v}^{k}$ and $\Omega_{j} \bar{v}^{j}$ in (\ref{eq:PertSingle1}). Thus we get
\be
(\bar{v}_j k^j) \left[ \left(\omega - \frac{c_s^2 k^2}{\omega}\right) + 4 \frac{\Omega^{2}}{\omega} \left(1 -  \cos^2{\theta} \frac{k^2 c_s^{2}}{\omega^2} \right) \right] = 0 \ .
\ee
Provided that the wave is not purely transverse, in which case we would have $\bar{v}_j k^j=0$,
we arrive at the dispersion relation
\be
\omega^4 - \omega^2 \left[ c_s^2 k^2 + 4 \Omega^{2} \right] + 4 \Omega^{2} \cos^2{\theta} k^2 c_s^{2} = 0 \ .
\label{disper}\ee
In the limit of slow  rotation, this quartic in $\omega$ has approximate solutions,
\bear
  \omega & \approx & \pm c_s k \left( 1 + \frac{2\Omega^2}{c_s^2 k^2} \sin^2{\theta} \right)  \approx  \pm c_s k \ , \\
  \omega & \approx & \pm 2 \Omega \cos{\theta} \ .
\eear
These are the well known results for sound waves and inertial waves, respectively.

Before we proceed, we need to consider whether the system may admit purely transverse waves.
This is important since (\ref{disper}) does not apply when $\bar{v}_j k^j=0$.
In this case, we first of all see from (\ref{eq:PertSingle3}) that we must also have $\Omega^i$ parallel with the 
wave vector $k^i$. The corresponding dispersion relation follows
 easily by taking the cross product of the Euler equation
(\ref{eq:PurtEqnSingle}) with $\Omega^i$.  This leads to
\be
\left( \omega - { 4 \Omega^2 \over \omega} \right) \bar{v}_i = 0 \ .
\ee
Hence we have $\omega = \pm 2 \Omega$. In other words, we obtain the 
$\theta\to 0$ limit for the inertial waves. Thus we learn that, in this limit
the inertial waves are purely transverse. By returning to the perturbed Euler 
equations, and representing the solution in a Cartesian coordinate system where
the $z$-axis is aligned with the wave vector, we see that we must have
$\bar{v}_x=\pm i \bar{v}_y$. This solution obviously satisfies $\bar{v}_j \bar{v}^j=0$.

\section{The two-fluid problem}

Let us now extend the plane-wave analysis to the two fluid problem. In principle, we expect
a doubling of the number of solutions. In addition, we are interested in the
new features that become relevant when we are dealing with a
multi-fluid situation. Our focus will be on the entrainment, which represents a non-dissipative coupling between the fluids
and the mutual friction, which represents dissipation due to electrons scattering off of the neutron vortices.

As we want to account for vortex effects it is natural to consider the
equations of motion in a rotating frame.  We will assume that the
background equilibrium is such that the two fluids rotate
together. This situation, which would correspond to the two fluids being in chemical equilibrium, is
slightly simplified which  keeps the analysis manageable.
Having said that, it is worth emphasising that we also need to
understand what happens when there is a velocity difference in the background. Since
relative rotation is required for the standard explanation of the pulsar glitches,
one would in principle expect the two fluids in a neutron star core to rotate at different rates.
If we are primarily interested in dynamics on a short timescale compared to the
time it takes to re-establish chemical equilibrium we have the freedom to consider
different background velocities. This general problem is, however, quite complicated and it makes sense
to first consider the case where the two fluids rotate
together.

As discussed by, for example \citet{PRIX,NILSCOMER}, the required equations of motion can be written 
(in a frame rotating uniformly with angular velocity $\Omega^i$)
\bear
\left( \pd{}{t} + v^{j}_{\X} \nabla_{j} \right) \left( v_{i}^{\X} + \varepsilon_{\X} w_{i}^{\Y\X} \right) + \varepsilon_{\X} w_{j}^{\Y\X}  \nabla_{i} v_{\X}^{j} + \nabla_{i} (\Phi + \tilde{\mu}_{\X}) + 2\epsilon_{ijk} \Omega^{j} v^{k}_{\X} = f_i^\X \ .
\eear
The constituent indices $\X$ and $\Y$ ($\X \neq \Y$) label the fluids, and will later take the values $\n$ and $\p$.
 The former
represents the superfluid neutrons and the latter a charge neutral conglomerate of protons and electrons.
The relative velocity is denoted by $w_i^{\X\Y} = v_i^\X-v_i^\Y$. 
The force  $f_i^\X$ on the right-hand side  represents a dynamical coupling between the two fluids. We will focus on the case when
it arises due to electron scattering off of the magnetic fields associated with the rotational vortices in the neutron superfluid.
Then we have, assuming that the vortex array is straight \citep{ASC1},
\be
f_i^\X = \frac{\rho_{\n}}{\rho_{\X}} n_v \mathcal{B}^\prime \epsilon_{ijk} \kappa^{j} w_{\X\Y}^{k}  + \frac{\rho_{\n}}{\rho_{\X}} n_v \mathcal{B} \epsilon_{ijk}\hat{ \kappa}^{j} \epsilon^{klm} \kappa_{l} w^{\X\Y}_{m} \ .
\label{fmf}\ee
In this expression $n_v$ is the vortex number density per unit area, and at
 the macroscopic level we have \citep{ASC2}
\be
n_v \kappa_i = \epsilon_{ijk} \nabla^j ( v_\n^k + \varepsilon_\n w_{\p\n}^k) \ .
\label{momdef}\ee
This means that for a straight vortex array, representing bulk rotation, we would have 
\be
n_v \kappa_{i} = 2 \Omega_{i}^{\n} + 2 \varepsilon_{\n} ( \Omega_{i}^{\p} - \Omega_{i}^{\n}) \ .
\ee
In our chosen background configuration the two fluids rotate together, so the second term in the above expression
vanishes. 
 As usual, mass conservation requires that
\be
\pd{\rho_{\X}}{t} + \nabla_{j}(\rho_{\X} v^{j}_{\X}) = 0 \ .
\ee

We consider small perturbations away from a background where the two fluids are at rest in the rotating frame. That is, we use
\be
\pd{\delta \rho_{\X}}{t} + \nabla^{j}(\rho_{\X} \delta v_{j}^{\X}) = 0 \ .
\ee
Making also the Cowling approximation ($ \delta \Phi = 0$),  the perturbed equations of motion become
\bear
\pd{}{t} \left( \delta v_{i}^{\X} + \varepsilon_{\X} \delta w_{i}^{\Y\X} \right) + \nabla_{i} \delta \tilde{\mu}_{\X} + 2\epsilon_{ijk} \Omega^{j} \delta v^{k}_{\X} =
\delta f_i^\X \ ,
\eear
where
\be
\delta f_i^\X = \frac{\rho_{\n}}{\rho_{\X}} n_v \mathcal{B}^\prime \epsilon_{ijk} \kappa^{j} \delta w_{\X\Y}^{k}  + \frac{\rho_{\n}}{\rho_{\X}}
n_v \mathcal{B}
\epsilon_{ijk}\hat{\kappa}^{j} \epsilon^{klm} \kappa_{l} \delta w^{\X\Y}_{m} \ .
\ee
It should be noted that setting the background rotation rates equal greatly reduces the complexity of the perturbed mutual friction force.

We next assume that the perturbations can be represented by plane waves. Then we have
\bear
\delta v_{i}^{\X} = \bar{v}_{i}^{\X} e^{i(\omega t +  k_{j} x^{j})} \ , \non
\delta \rho_{\X} = \bar{\rho}_{\X} e^{i(\omega t + k_{j} x^{j} )} \ ,
\eear
where $\bar{v}_{i}^{\X}$ and $\bar{\rho}_{\X}$ are assumed to be constant (that is, they vary on lengthscales much longer than the wavelength).
The continuity equations then become
\bear \label{eq:PerturContEqn}
  k_{j} \bar{v}^{j}_{\X} = - \frac{\bar{\rho}_{\X}}{\rho_{\X}} \omega \ .
\eear

To make progress we need a representation of the equation of state. In our formalism \citep{PRIX,NILSCOMER}, we need to provide an energy functional $E$
from which the chemical potentials follow according to
\be
\tilde{\mu}_\X = \left. { \partial E \over \partial \rho_\X} \right|_{\rho_\Y, w_{\Y\X}^2} \ .
\ee
The entrainment $\varepsilon_\X$ is similarly determined as
\be
\varepsilon_\X = { 2 \alpha \over \rho_\X} \ , \mbox{ where } \quad \alpha = \left. { \partial E \over \partial w_{\Y\X}^2} \right|_{\rho_\X, \rho_\Y} \ .
\ee
The basic idea is that if the system is isotropic, then the energy function can be constructed from the various scalars that can be formed from
the dynamical variables $\rho_\X$ and $v_\X^i$.
This means that one would generally expect to have
\be
\delta \tilde{\mu}_{\X} = \pd{\tilde{\mu}_{\X}}{\rho_{\X}} \delta \rho_{\X} + \pd{ \tilde{\mu}_{\X}}{\rho_{\Y}} \delta \rho_{\Y} +
 \pd{ \tilde{\mu}_{\X}}{w_{\Y\X}^2} \delta w_{\Y\X}^2 \ .
\ee
In the present case, where the two fluids move together in the background, 
the last term will vanish since  $\delta w_{\Y\X}^2 = 2 w_{\Y\X}^j \delta w^{\Y\X}_j=0$.
Hence, we can use
\be
  \delta \tilde{\mu}_{\X} = \pd{\tilde{\mu}_{\X}}{\rho_{\X}} \delta \rho_{\X} + \pd{ \tilde{\mu}_{\X}}{\rho_{\Y}} \delta \rho_{\Y}
  = \tilde{\mu}_{\X\X} \delta \rho_{\X} + \tilde{\mu}_{\X\Y} \delta \rho_{\Y} \ .
\ee
Later, when we allow for relative flow in the background, this form for the perturbed chemical potentials will no longer be generally valid.
We will nevertheless use it in order to simplify the analysis. In practice, this amounts to assuming that the
equation of state is ``separable'' in the sense that it can be written\footnote{It is worth noting that if one wanted to 
propose a ``realistic'' separable equation of state, then  one would have to think more carefully about 
the units of the two terms. In other words, our proposed form for $E$ should be seen as a mathematical construction motivated by the 
fact that it simplifies the analysis.}
\be
E = f(\rho_\n, \rho_\p) + g(w_{\Y\X}^2) \ .
\ee

Under the above assumptions, the momentum equations become
\bear
  i \omega( \bar{v}_{i}^{\X} + \varepsilon_{\X} \bar{w}^{\Y\X}_{i}) - i k_{i} (\tilde{\mu}_{\X\X} \bar{\rho}_{\X} + \tilde{\mu}_{\X\Y} \bar{\rho}_{\Y}) + 2 \epsilon_{ijk} \Omega^{j} \bar{v}_{\X}^{k} = 2\frac{\rho_{\n}}{\rho_{\X}} \mathcal{B}' \epsilon_{ijk} \Omega^{j}\bar{w}_{\X\Y}^{k} + 
2\frac{\rho_{\n}}{\rho_{\X}} \mathcal{B} \epsilon_{ijk} \hat{\kappa}^{j} \epsilon^{klm} \Omega_{l} \bar{w}^{\X\Y}_{m} \ .
\eear
 After using the perturbed continuity equation (\ref{eq:PerturContEqn}) and rearranging we arrive at
\bear
  \bar{v}^{\X}_{m} \left[ i \omega (1-\varepsilon_{\X}) \delta^{m}_{i} + 2 \epsilon_{ij} {}^{m} \Omega^{j} - 2 \frac{\rho_{\n}}{\rho_{\X}} \mathcal{B}' \epsilon_{ij} {}^{m} \Omega^{j} - 2 \frac{\rho_{\n}}{\rho_{\X}} \mathcal{B} \epsilon_{ijk} \hat{\kappa}^{j} \epsilon^{klm} \Omega_{l} - i k_{i} \tilde{\mu}_{\X\X} \frac{\rho_{\X}}{\omega} k^{m} \right] \non
  + \bar{v}_{m}^{\Y} \left[ i \omega \varepsilon_{\X} \delta_{i}^{m} + 2 \frac{\rho_{\n}}{\rho_{\X}} \mathcal{B}' \epsilon_{ij} {}^{m} \Omega^{j} + 2 \frac{\rho_{\n}}{\rho_{\X}} \mathcal{B} \epsilon_{ijk} \hat{\kappa}^{j} \epsilon^{klm} \Omega_{l} -  i k_{i} \tilde{\mu}_{\X\Y} \frac{\rho_{\Y}}{\omega} k^{m} \right] = 0 \ .
\eear
Let us now introduce the speeds of sound as \citep{mnras}
\be
c_\X^2 = \rho_\X \tilde{\mu}_{\X\X} \ ,
\label{c_sound}\ee
and represent the ``chemical coupling'' by
\be
\mathcal{C}_\X = \rho_\X \tilde{\mu}_{\X\Y} = \rho_\X \tilde{\mu}_{\Y\X} \ .
\ee
Since the partial derivatives commute we have
\be
\mathcal{C}_\p = { \rho_\p \over \rho_\n} \mathcal{C}_\n \ .
\ee
Then
we have
\bear
\label{eq:PerturbedNSwSV}
  \bar{v}^{\X}_{m} \left[ i \omega (1-\varepsilon_{\X}) \delta^{m}_{i} + 2 \epsilon_{ij} {}^{m} \Omega^{j} - 2 \frac{\rho_{\n}}{\rho_{\X}} \mathcal{B}' \epsilon_{ij} {}^{m} \Omega^{j}  - 2 \frac{\rho_{\n}}{\rho_{\X}} \mathcal{B} \epsilon_{ijk} \hat{\kappa}^{j} \epsilon^{klm} \Omega_{l}  - i k_{i}\frac{c_{\X}^2}{\omega} k^{m} \right]  \non
  + \bar{v}_{m}^{\Y} \left[ i \omega \varepsilon_{\X} \delta_{i}^{m} + 2 \frac{\rho_{\n}}{\rho_{\X}} \mathcal{B}' \epsilon_{ij} {}^{m} \Omega^{j}  + 2 \frac{\rho_{\n}}{\rho_{\X}} \mathcal{B} \epsilon_{ijk} \hat{\kappa}^{j} \epsilon^{klm} \Omega_{l}  -  i k_{i} \frac{\mathcal{C}_{\Y}}{\omega} k^{m} 
\right] = 0 \ .
\eear
The dispersion relation for wave propagation is encoded in this equation. As in the single fluid case,
 it is preferable to work with scalar equations.
Hence, we contract  (\ref{eq:PerturbedNSwSV}) with $k_{i}$. This gives
\bear \label{eq:KdotEuler}
\bar{v}^{\X}_{m} \left[ i \omega (1-\varepsilon_{\X}) k^{m} + \left( 2 - 2 \frac{\rho_{\n}}{\rho_{\X}} \mathcal{B}' \right) k^{i} \epsilon_{ij} {}^{m} \Omega^{j}  - 2 k^{i} \frac{\rho_{\n}}{\rho_{\X}} \mathcal{B} \epsilon_{ijk} \hat{\kappa}^{j} \epsilon^{klm} \Omega_{l}  - i k^{2}  \frac{c_{\X}^2}{\omega} k^{m} \right] \non
+ \bar{v}_{m}^{\Y} \left[ i \omega \varepsilon_{\X} k^{m} + 2 \frac{\rho_{\n}}{\rho_{\X}} \mathcal{B}' k^{i} \epsilon_{ij} {}^{m} \Omega^{j}  + 2 \frac{\rho_{\n}}{\rho_{\X}} \mathcal{B} k^{i} \epsilon_{ijk} \hat{\kappa}^{j} \epsilon^{klm} \Omega_{l}  -  i k^{2}  \frac{\mathcal{C}_{\Y}}{\omega} k^{m} \right] = 0 \ .
\eear
We obtain a second scalar equation by contracting (\ref{eq:PerturbedNSwSV}) with $\Omega_{i} \epsilon^{ijk} k_{j}$;
\bear \label{eq:KcrossEuler}
\bar{v}^{\X}_{m} \left[ i \omega ( 1 - \varepsilon_{\X} ) \Omega_{i} \epsilon^{ikm} k_{k} + \left( 2 - 2 \frac{\rho_{\n}}{\rho_{\X}} \mathcal{B}' \right) \Omega_{i} \epsilon^{ikl} k_{k} \epsilon_{lj} {}^{m} \Omega^{j}  - 2 \frac{\rho_{\n}}{\rho_{\X}} \mathcal{B} \Omega_{i} \epsilon^{iqr} k_{q} \epsilon_{rjk} \hat{\kappa}^{j} \epsilon^{klm} \Omega_{l}  \right] \non
+ \bar{v}_{m}^{\Y} \left[ i \omega \varepsilon_{\X} \Omega_{i} \epsilon^{ijm} k_{j} + 2 \frac{\rho_{\n}}{\rho_{\X}} \mathcal{B}' \Omega_{i} \epsilon^{ikl} k_{k} \epsilon_{lj} {}^{m} \Omega^{j}  + 2 \frac{\rho_{\n}}{\rho_{\X}} \mathcal{B} \Omega_{i} \epsilon^{iqr} k_{q} \epsilon_{rjk} \hat{\kappa}^{j} \epsilon^{klm} \Omega_{l}  \right] = 0 \ .
\eear
A third equation follows from the contraction of  (\ref{eq:PerturbedNSwSV}) with $\Omega^{i}$;
\bear
\label{eq:OmegaDotEuler}
\bar{v}^{\X}_{m} \left[ i \omega (1-\varepsilon_{\X}) \Omega^{m} - i k_{i} \Omega^{i}   \frac{c_{\X}^2}{\omega} k^{m} \right]
+ \bar{v}_{m}^{\Y} \left[ i \omega \varepsilon_{\X} \Omega^{m} - i k_{i} \Omega^{i}  \frac{\mathcal{C}_{\Y}}{\omega} k^{m} \right] = 0 \ .
\eear

Although in principle straightforward, the formulation of the dispersion relation is still rather messy.
To facilitate its construction it is useful to introduce some further notation. To motivate this, let us
write out the two equations (\ref{eq:OmegaDotEuler}) explicitly. We have
\bear
\bar{v}^{\n}_{m} \left[ i \omega (1-\varepsilon_{\n}) \Omega^{m} - i k_{i} \Omega^{i}   \frac{c_{\n}^2}{\omega} k^{m} \right]
+ \bar{v}_{m}^{\p} \left[ i \omega \varepsilon_{\n} \Omega^{m} - i k_{i} \Omega^{i}  \frac{\mathcal{C}_{\p}}{\omega} k^{m} \right] = 0 \ ,
\eear
\bear
\bar{v}^{\p}_{m} \left[ i \omega (1-\varepsilon_{\p}) \Omega^{m} - i k_{i} \Omega^{i}  \frac{c_{\p}^2}{\omega} k^{m} \right]
+ \bar{v}_{m}^{\n} \left[ i \omega \varepsilon_{\p} \Omega^{m} - i k_{i} \Omega^{i}  \frac{\mathcal{C}_{\n}}{\omega} k^{m} \right] = 0 \ ,
\eear
If we define
\be
\vc{\xi} = \Mat{1-\varepsilon_{\n}}{\varepsilon_{\n}}{\varepsilon_{\p}}{1-\varepsilon_{\p}} \ , \qquad \vc{\tilde{\mu}} = \Mat{ \c_{\n}^2}{ \mathcal{C}_{\p}}{\mathcal{C}_{\n}}{c_{\p}^2} \ , \qquad  \vc{v}_{\Omega} = \Vec{\Omega^{j} \bar{v}_{j}^{\n}}{\Omega^{j} \bar{v}_{j}^{\p}} \ , \qquad \vc{v}_{k} = \Vec{k^{j} \bar{v}_{j}^{\n}}{k^{j} \bar{v}_{j}^{\p}} \ ,
\ee
then (\ref{eq:OmegaDotEuler}) can be rewritten as
\be
\label{eq:Algebra1}
\frac{\omega^{2}}{k_{i}\Omega^{i} } \vc{\xi} \vc{v}_{\Omega} = \vc{\tilde{\mu}} \vc{v}_{k} \ .
\ee
As a note for future reference, we will also use
\be
\vc{\xi}^{-1} = \frac{1}{\xi} \Mat{1-\varepsilon_{\p}}{-\varepsilon_{\n}}{-\varepsilon_{\p}}{1-\varepsilon_{\n}}
\ , \qquad \xi = \det \vc{\xi} = 1 - \varepsilon_{\n} - \varepsilon_{\p}
\ , \qquad \vc{v}_{\epsilon} = \Vec{\Omega^{i}  \epsilon_{ijk} k^{j} \bar{v}^{k}_{\n}}{\Omega^{i}  \epsilon_{ijk} k^{j} \bar{v}^{k}_{\p}}
\ , \qquad \vc{\rho}  = \Mat{\rho_{\p}}{-\rho_{\p}}{-\rho_{\n}}{\rho_{\n}} \ .
\ee

We can now rewrite (\ref{eq:KdotEuler}) and (\ref{eq:KcrossEuler}) as
\be
\label{eq:Algebra2}
\left( i \omega \vc{\xi} - i \frac{k^2}{\omega} \vc{\tilde{\mu}} + 2 \frac{\mathcal{B}}{\rho_{\p}}\Omega \vc{\rho}_0  \right) \vc{v}_{k}
- \left( 2 \vc{I} - 2 \frac{\mathcal{B}'}{\rho_{\p}} \vc{\rho}_{0} \right) \vc{v}_{\epsilon}
- 2 \frac{\mathcal{B}}{\rho_{\p} \Omega}(k_j \Omega^j) \vc{\rho}_{0} \vc{v}_{\Omega} = 0 \ ,
\ee
and
\be
\label{eq:Algebra3}
\left( i \omega \vc{\xi} + 2 \frac{\mathcal{B} \Omega}{\rho_{\p}} \vc{\rho}_{0} \right) \vc{v}_{\epsilon}
+ \left( 2 \vc{I} - 2 \frac{\mathcal{B}'}{\rho_{\p}} \vc{\rho}_{0} \right)\Omega^{2} \vc{v}_{k}
- \left( 2 \vc{I} - 2 \frac{\mathcal{B}'}{\rho_{\p}} \vc{\rho}_{0} \right) (k_j \Omega^j) \vc{v}_{\Omega} = 0 \ .
\ee

Assuming that $1 - \varepsilon_{\n} - \varepsilon_{\p} \neq 0$,
and substituting (\ref{eq:Algebra1}) into (\ref{eq:Algebra2}) and (\ref{eq:Algebra3}) we get
\be \label{eq:MatEul1}
\left[ i \omega \vc{\xi} - i \frac{k^2}{\omega} \vc{\tilde{\mu}} + 2 \frac{\mathcal{B}}{\rho_{\p} \Omega}\vc{\rho}_{0} \left( \Omega^{2} - \frac{(k_j \Omega^j)^{2}}{\omega^{2}} \vc{\xi}^{-1} \vc{\tilde{\mu}} \right) \right] \vc{v}_{k}
- \left( 2 \vc{I} - 2 \frac{\mathcal{B}'}{\rho_{\p}} \vc{\rho}_{0} \right) \vc{v}_{\epsilon} = 0 \ ,
\ee
and
\be \label{eq:MatEul2}
\left( i \omega \vc{\xi} + 2 \frac{\mathcal{B} \Omega}{\rho_{\p}} \vc{\rho}_{0} \right) \vc{v}_{\epsilon}
= - \left[ \left(2 \vc{I} - 2 \frac{\mathcal{B}'}{\rho_{\p}} \vc{\rho}_{0}\right) \left( \Omega^{2}  -  \frac{(k_j \Omega^j)^{2}}{\omega^{2}} \vc{\xi}^{-1} \vc{\tilde{\mu}} \right) \right] \vc{v}_{k} \ .
\ee
We now define
\be
\vc{\rho}_{0}^\mathrm{zero} = \Mat{\rho_{\n}}{\rho_{\p}}{\rho_{\n}}{\rho_{\p}} \ ,
\ee
such that $\vc{\rho}_{0}^\mathrm{zero} \vc{\rho}_{0} = \vc{0}$.
The matrix on the left hand side of equation (\ref{eq:MatEul2}) then has the inverse,
\be
\left[ - \omega^{2} \xi + 2 i \omega \Omega \mathcal{B} \left(1 + \frac{\rho_{\n}}{\rho_{\p}} \right) \right]^{-1}
\left( i \omega \xi \vc{\xi}^{-1} + 2 \frac{\mathcal{B} \Omega}{\rho_{\p}} \vc{\rho}_{0}^\mathrm{zero} \right) \ .
\ee
Using this in (\ref{eq:MatEul1}) we obtain
\bear
\lefteqn{\left[ i \omega \vc{\xi} - i \frac{k^2}{\omega} \vc{\tilde{\mu}} + 2 \frac{\mathcal{B}}{\rho_{\p} \Omega}\vc{\rho}_{0} \left( \Omega^{2} - \frac{(k_j \Omega^j)^{2}}{\omega^{2}} \vc{\xi}^{-1} \vc{\tilde{\mu}} \right) \right] \vc{v}_{k}} \non
& & {} - \left( 2 \vc{I} - 2 \frac{\mathcal{B}'}{\rho_{\p}} \vc{\rho}_{0} \right) \frac{i \omega \xi \vc{\xi}^{-1} + 2 \frac{\mathcal{B} \Omega}{\rho_{\p}} \vc{\rho}_{0}^\mathrm{zero}}{ \omega^{2} \xi - 2 i \omega \Omega \mathcal{B} \left(1 + \frac{\rho_{\n}}{\rho_{\p}} \right)}
\left[ \left( 2 \vc{I} - 2 \frac{\mathcal{B}'}{\rho_{\p}} \vc{\rho}_{0} \right) \left( \Omega^{2}  -  \frac{(k_j \Omega^j)^{2}}{\omega^{2}} \vc{\xi}^{-1} \vc{\tilde{\mu}} \right) \right] \vc{v}_{k} = 0 \ .
\eear
By using $\vc{\rho}_{0} \vc{\rho}_{0}^\mathrm{zero} = \vc{0} $ this expands to give
\bear
\lefteqn{\vc{v}_{k}  \left\{ i \omega \vc{\xi} - i \frac{k^2}{\omega} \vc{\tilde{\mu}} + 2 \frac{\mathcal{B}}{\rho_{\p} \Omega}\vc{\rho}_{0} \left( \Omega^{2} - \frac{(k_j \Omega^j)^{2}}{\omega^{2}} \vc{\xi}^{-1} \vc{\tilde{\mu}} \right)
 - \frac{ \Omega^{2}}{ \omega^{2} \xi - 2 i \omega \Omega \mathcal{B} \left(1 + \frac{\rho_{\n}}{\rho_{\p}} \right)} \right.} \non
& & \left< 4 i \omega \xi \vc{\xi}^{-1} - 4 i \omega \xi \frac{\mathcal{B}'}{\rho_{\p}} [ \vc{\rho}_{0} \vc{\xi}^{-1} + \vc{\xi}^{-1} \vc{\rho}_{0} ] + i \omega \xi \left( \frac{ 2 \mathcal{B}'}{\rho_{\p}} \right)^{2} \vc{\rho}_{0} \vc{\xi}^{-1} \vc{\rho}_{0}
+ 8 \frac{\mathcal{B} \Omega}{\rho_{\p}} \vc{\rho}_{0}^\mathrm{zero}  - \frac{(k_j \Omega^j)^{2}}{\omega^{2} \Omega^{2}} \right. \non
& & \left. \left.
\left[ 4 i \omega \vc{\xi}^{-1} \vc{\xi}^{-1} \vc{\tilde{\mu}} - 4 i \omega \frac{\mathcal{B}'}{\rho_{\p}} ( \vc{\rho}_{0} \vc{\xi}^{-1} + \vc{\xi}^{-1} \vc{\rho}_{0} ) \vc{\xi}^{-1} \vc{\tilde{\mu}} + i \omega \left( \frac{2 \mathcal{B}'}{\rho_{\p}} \right)^{2} \vc{\rho}_{0} \vc{\xi}^{-1} \vc{\rho}_{0} \vc{\xi}^{-1} \vc{\tilde{\mu}} +  8 \frac{\mathcal{B} \Omega}{\rho_{\p} \zeta} \vc{\rho}_{0}^\mathrm{zero} \vc{\xi}^{-1} \vc{\tilde{\mu}} \right] \right> \right\}  = 0 \ .
\eear
In order for us to have $\vc{v}_{k} \neq 0$ the determinant of the matrix in the curly brackets must vanish.
This condition provides the dispersion relation for waves in the two-fluid system.

As in the single-fluid problem, our analysis does not apply to waves that are purely transverse.
Such waves are, however, not very likely unless we align the wave vector with the rotation axis. 
In the general case, we see from (\ref{eq:MatEul1}) and (\ref{eq:MatEul2}) that
we can have purely transverse waves (for which $\vc{v}_k=0$) only if also $\vc{v}_\epsilon=0$ or if
\bear
\det{ \left(i \omega \vc{\xi} + \frac{\mathcal{B} \Omega}{\rho_{\p}} \vc{\rho}_{0} \right) } & = & 0 \ , \non
\det{ \left( 2 \vc{I} - \frac{\mathcal{B}'}{\rho_{\p}} \vc{\rho}_{0} \right) } & = & 0 \ .
\eear
These conditions lead to
\bear
0 & = & \omega^{2} (1- \varepsilon_{\n} - \varepsilon_{\p}) - i \omega \Omega \mathcal{B} \left( 1 + \frac{\rho_{\n}}{\rho_{\p}} \right) \ , \non
0 & = & 4 - 2 \mathcal{B}' \left( 1 + \frac{\rho_{\n}}{\rho_{\p}} \right) \ . 
\eear
The second condition is extremely  restrictive and so a purely transverse wave is unlikely. In fact, for typical neutron star conditions
we expect $\mathcal{B}'\ll 1$ \citep{ASC1} which suggests that purely transverse waves are not possible unless the wave is aligned with the rotation
in such a way that $\vc{v}_\epsilon=0$. Although somewhat contrived, this particular case is  interesting and we will discuss 
it in more detail in Sections~5 and 6.

\section{Illustrative examples}

In the previous section we wrote down all the relations we need to derive the general
dispersion relation for the two-fluid problem. It should be clear
that, since the generic dispersion relation is a high order polynomial in
$\omega$, this problem is quite rich. In order to understand the solutions it is useful to
consider a sequence of increasingly complex model situations.
This will give us a feeling for how the various parameters in the model affect the wave
propagation.

\subsection{No rotation, coupling, or friction}

It is natural to start with the very simplest case,
with the two fluids completely decoupled.
This model corresponds to an equation of state of form
\be
E = f(n_{\n}) + g(n_{\p}) \ .
\label{separate}\ee
This leads to $\tilde{\mu}_{\X\Y} = 0$ and $\varepsilon_\X=0$. In essence, the
two fluids are not coupled either chemically or by entrainment. If we also assume that
there is no background rotation or friction in the system, i.e. let
 $\Omega = \mathcal{B} = \mathcal{B}'=0$, then the dispersion relation follows from
 the determinant
\be
\left| i \omega \vc{I} - i \frac{k^2}{\omega} \vc{\tilde{\mu}} \right| = 0 \ .
\ee
This expands to
\be
\omega^2 \left\{ 1 - \frac{k^2}{\omega^2} c_\n^2 \right\} \left\{ 1 - \frac{k^2}{\omega^2} c_\p^2\right\} = 0 \ .
\ee
which has the non-trivial solutions 
\be
\omega^2 = k^2 c_\n^2 \qquad \mbox{ and } \qquad  \omega^2 = k^2 c_\p^2 \ .
\ee
Hence, we have the anticipated result that the system only supports sound waves,
\bear
\omega & = & \pm k c_{\n} \ , \non
\omega & = & \pm k c_{\p} \ .
\eear
It is also easy to show that these waves are longitudinal, as one would expect.

\subsection{Including entrainment}

We can now begin to investigate how various coupling mechanisms modify these waves.
Let us first consider the entrainment. Then we need an equation of state that depends on
the relative velocity. Thus, we assume that
\be
E = f(n_{\n}) + g(n_{\p}) + h(w_{\n\p}^{2}) \ .
\ee
This is obviously not the general case, but since we want to be able to
analyze the problem analytically it is natural to restrict ourselves to
this class of separable models. It is straightforward to study more
generic situations numerically, but the results should not differ
qualitatively from the ones we discuss here.

Still assuming that $\Omega = \mathcal{B} = \mathcal{B}' = \tilde{\mu}_{\X\Y} = 0$ we obtain
the dispersion relation from
\be
\left| i \omega \vc{\xi} - i \frac{k^2}{\omega} \vc{\tilde{\mu}} \right| = 0 \ .
\ee
We expand this to get
\be
\omega^2 \left\{ \left( 1 - \varepsilon_{\n} -  \frac{k^2}{\omega^2} c_{\n}^{2} \right) \left( 1 - \varepsilon_{\p} - \frac{k^2}{\omega^2} c_\p^2 \right) - \varepsilon_{\n} \varepsilon_{\p} \right\} = 0 \ .
\label{entdisp}\ee
%Here it is useful to recall that
%\be
%\rho_\X \varepsilon_\X = 2 \alpha
%\ee
The solutions are found from
\be
\omega^{4} \xi - \omega^{2} \left[ (1 - \varepsilon_{\n}) k^2 c_{\p}^{2} + (1 - \varepsilon_{\p}) k^2 c_{\n}^{2} \right] + k^4 c_{\n}^{2} c_{\p}^{2} = 0
\ .
\ee
Solving for $\omega^2$ we have
\be
\omega^2 = \frac{1}{2 \xi} \left[ (1 - \varepsilon_{\n}) k^2 c_{\p}^{2} + (1 - \varepsilon_{\p}) k^2 c_{\n}^{2} \right] \pm
{ k^2 \over 2 \xi} \left\{  \left[ (1 - \varepsilon_{\n}) c_{\p}^{2} + (1 - \varepsilon_{\p}) c_{\n}^{2} \right]^2 - 4 \xi c_{\n}^{2} c_{\p}^{2} \right\}^{1/2} \ .
\ee
To make further progress it is useful to assume that the entrainment is a small effect and use Taylor expansion in $\varepsilon_{\X}$.
We see immediately from (\ref{entdisp}) that  to linear order in entrainment
the frequencies are  given by
\bear \label{eq:EntrainCorrecSound}
\omega & \approx & \pm (1 + \frac{1}{2} \varepsilon_{\n}) k c_{\n} \ , \non
\omega & \approx & \pm (1 + \frac{1}{2} \varepsilon_{\p}) k c_{\p} \ .
\eear
This illustrates how the sound waves are affected by a weak entrainment coupling.
At this level there appears to be no interaction between the two  wave speeds.
This would be a higher order effect for this equation of state.

\subsection{Chemical coupling}

Let us  consider the other way that the two fluids in a non-rotating system may be coupled.
In order to see what effect chemical coupling has on the waves, we consider an
equation of state of form
\begin{equation}
  E = f(n_{\n},n_{\p}) \ .
\end{equation}
The key difference is that the chemical potential of one fluid can now
be affected by the population density of the other constituent.
This requires us to work out
\begin{equation}
  \left| \omega \vc{\xi} - \frac{k^{2}}{\omega} \tilde{\vc{\mu}} \right| = 0 \ ,
\end{equation}
which expands to give
\be
\omega^{4} - \omega^{2}  k^2 (c_{\n}^{2} + c_{\p}^{2}) + k^4 \left( c_{\n}^{2} c_{\p}^{2} - \frac{\rho_{\p}}{\rho_{\n}} \mathcal{C}^2_\n
\right) \ .
\ee
Taking $\mathcal{C}^{2}_\n$ as small and solving for $\omega$, we get either
\begin{equation}
  \omega = \pm k c_{\n} \left[ 1 + \frac{\rho_{\p}}{2 \rho_{\n}} \frac{\mathcal{C}^{2}_\n}{k^2 c_{\n}^2( c_{\n}^{2} - c_{\p}^{2} )} \right] \ ,
\end{equation}
or
\begin{equation}
  \omega = \pm k c_{\p} \left[ 1 + \frac{\rho_{\p}}{2 \rho_{\n}}  \frac{\mathcal{C}^{2}_\n}{k^2 c_{\p}^2( c_{\p}^{2} - c_{\n}^{2} )} \right] \ .
\end{equation}
These are still modified sound waves associated with each constituent.

\subsection{Slow rotation}

We now move on to the case of slow rotation.
In addition to the sound waves, we expect to find inertial modes. Since we are assuming that the
two fluids co-rotate in the background the inertial modes are likely to be degenerate.
To keep the problem simple, we  assume that there is no chemical coupling.  In practice, we again
let the equation of state be of the form (\ref{separate}).
Then taking $\mathcal{B} = \mathcal{B}' = \varepsilon_{\X} = \tilde{\mu}_{\X\Y} = 0$ the
dispersion relation follows from
\be
\left| i \omega \vc{I} - i \frac{k^2}{\omega} \vc{\tilde{\mu}} - \frac{\Omega^2}{\omega^2} \left( 4 i \omega \vc{I} - 4 i \frac{k^2}{\omega}\cos^2{\theta} \vc{\tilde{\mu}} \right) \right| = 0 \ .
\ee
As before,  ${\theta}$ is the angle between $k_{i}$ and $\Omega_{i} $ such that $k^{j} \Omega_{j}  = k \Omega \cos{\theta}$.
The determinant expands to give
\be
\omega^{2} \left\{ 1 - \frac{k^2}{\omega^2} c_{\n}^{2} - \frac{4 \Omega^{2}}{\omega^2} + 4 \frac{\Omega^{2} k^{2}}{\omega^{4}} \cos^2{\theta} c_{\n}^{2} \right\} \left\{ 1 - \frac{k^2}{\omega^2} c_{\p}^{2} - \frac{4 \Omega^{2}}{\omega^2} + 4 \frac{\Omega^{2} k^{2}}{\omega^{4}} \cos^2{\theta} c_{\p}^{2} \right\} = 0 \ .
\ee
There are clearly two decoupled cases.
The two sets of  solutions are found from
\bear
\label{eq:QuadSrNcNfN}
0 & = & \omega^{4} - \omega^{2} \left[ k^2 c_{\n}^{2} + 4 \Omega^{2} \right] + 4 \Omega^{2} k^{2} \cos^2{\theta} c_{\n}^{2} \ , \\
\label{eq:QuadSrNcNfP}
0 & = & \omega^{4} - \omega^{2} \left[ k^2 c_{\p}^{2} + 4 \Omega^{2} \right] + 4 \Omega^{2} k^{2}\cos^2{\theta} c_{\p}^{2} \ ,
\eear
If we for simplicity assume slow rotation,  the solutions to (\ref{eq:QuadSrNcNfN}) are
\bear
\omega & \approx & \pm  k c_{\n} \left(1 + { 2 \Omega^{2}\over k^2 c_\n^2}  \sin^{2}{\theta} \right) \ , \label{eq:RotModeSoundNeut} \\
\omega & \approx & \pm 2 \Omega \cos{\theta} \label{eq:InertialModeNeutron} \ ,
\eear
while
the solutions to (\ref{eq:QuadSrNcNfP}) are,
\bear
\omega & \approx & \pm  k c_{\p}  \left(1 + {2 \Omega^{2}  \over k^2 c_{\p}^2 } \sin^{2}{\theta} \right) \ , \label{eq:RotModeSoundProt} \\
\omega & \approx & \pm 2 \Omega \cos{\theta} \ . \label{eq:InertialModeProton}
\eear
The solutions (\ref{eq:RotModeSoundNeut}) and (\ref{eq:RotModeSoundProt}) represent sound waves with a correction due to
the slow rotation. Solutions (\ref{eq:InertialModeNeutron}) and (\ref{eq:InertialModeProton}) are the (in this case
degenerate) inertial modes. The form of the solutions is exactly as one would expect from the single fluid problem.

\subsection{Mutual friction}

The simple cases we have considered so far provide an insight into the
different classes of waves that will be present in the
rotating two-fluid problem. We now want to develop an understanding of how these
waves are affected by the mutual friction. To do this, it is natural to assume that the
induced damping is weak. In the neutron star case, we also expect to have $\mathcal{B}'\ll \mathcal{B}$
which allows us to simplify the problem. \citet{ASC1} showed that $\mathcal{B}'= \mathcal{B}^2$, and the 
``canonical'' value for $\mathcal{B}$ is $\sim 4\times 10^{-4}$. 
 Since we are assuming that the mutual friction is weak, it is natural to
include it as a perturbation of the solutions we  found previously.

In order to be consistent we cannot consider the
effect of mutual friction without at the same time accounting for  rotation. Without
rotation there would be no neutron vortices in the background and hence no mutual
friction.
We therefore consider the situation when both $\mathcal{B}$ and $\Omega$ can be assumed small
(in a suitable sense). To make the analysis tractable we
assume that  $\varepsilon_{\X} = \tilde{\mu}_{\X\Y} = 0$.
Stricly speaking, it is not consistent to neglect the entrainment here. It plays a central role in
generating the mutual friction since it is the entrained protons flowing around a neutron vortex that generates the 
main component of the vortex magnetic field \citep{ASC1}. Hence, if we neglect the entrainment then we should not have 
the mutual friction either. Of course, the two contributions have completely different effects on the dynamics. 
As long as we are mainly interested in the dissipation the assumptions we make here should be 
acceptable.

The equation that we need to solve can be written
\bear
\label{eq:JustNeedALabel}
\lefteqn{ \left| i \omega \vc{I} - \frac{i k^2}{\omega} \vc{\tilde{\mu}} + \frac{2\mathcal{B}}{\rho_{\p} \Omega} \vc{\rho_{0}} \left( \Omega^{2} \vc{I} - \frac{(k_j \Omega^j)^2}{\omega^2} \vc{\tilde{\mu}} \right) \right.} \non
& & {} \left. - \frac{\Omega^{2}}{\omega^2 - 2 i \omega \Omega \mathcal{B} \left( 1 + \frac{\rho_{\p}}{\rho_{\n}} \right)} \left\{ i 8 \omega \vc{I} + 4 \frac{\mathcal{B} \Omega}{\rho_{\p}} \vc{\rho}_{0}^\mathrm{zero} - \frac{(k_j \Omega^j)^{2}}{\omega^2 \Omega^{2}} \left[ i 4 \omega \vc{\tilde{\mu}} + 8\frac{\mathcal{B} \Omega}{\rho_{\p}} \vc{\rho}_{0}^\mathrm{zero} \vc{\tilde{\mu}} \right] \right\} \right| = 0 \ .
\eear
Before substituting the appropriate solutions for $\omega_{0}$ in (\ref{eq:JustNeedALabel}) we note that, 
when written out in full,  we need the determinant of a matrix of form
\be
  \Mat{a}{\mathcal{B} b}{\mathcal{B} c}{d} \ .
\ee
This means that, to first order in $\mathcal{B}$, the dispersion relation is either
$a = 0$ or  $d = 0$.
Using this fact, and assuming that the frequency $\omega$ will be replaced by $\omega_{0} + \delta \omega$
where $\omega_0$ represents one of the undamped solutions from Section~4.4, we need to solve either
\bear \label{eq:RotFriN}
\lefteqn{ \left\{ i \omega_{0}(1 + \frac{\delta \omega}{\omega_{0}}) - \frac{i k^2}{\omega_{0}}(1 - \frac{\delta \omega}{\omega_{0}}) c_{\n}^{2} + 2 \frac{\mathcal{B}}{\Omega} \left( \Omega^{2} - \frac{(k_j \Omega^j)^2}{\omega_{0}^2} c_{\n}^{2} \right)  - \frac{\Omega^{2}}{\omega_{0}^2} \left( 1 - 2 \frac{\delta \omega}{\omega_{0}} + 2 i\frac{\Omega \mathcal{B}}{\omega_{0}} \left( 1 + \frac{\rho_{\n}}{\rho_{\p}} \right) \right) \right.} \non
& & {} \left. \left< i 4 \omega_{0} \left( 1 + \frac{\delta \omega}{\omega_{0}} \right) + 8 \frac{\mathcal{B} \Omega}{\rho_{\p}} \rho_{\n} - \frac{(k_j \Omega^j)^{2}}{\omega_{0}^2 \Omega^{2}} \left( 1 - 2 \frac{\delta \omega}{\omega_{0}} \right)\left[ i 4 \omega_{0} \left( 1 + \frac{\delta \omega}{\omega_{0}} \right) c_{\n}^{2} + 8 \frac{\mathcal{B} \Omega}{\rho_{\p}} \rho_{\n} c_{\n}^{2} \right] \right> \right\} =0 \ ,
\eear
or
\bear \label{eq:RotFriP}
\lefteqn{ \left\{ i \omega_{0}(1 + \frac{\delta \omega}{\omega_{0}}) - \frac{i k^2}{\omega_{0}}(1 - \frac{\delta \omega}{\omega_{0}}) c_{\p}^{2} + 2 \frac{\mathcal{B} \rho_{\n}}{\Omega \rho_{\p}} \left( \Omega^{2} - \frac{(k_j\Omega^j)^2}{\omega_{0}^2} c_{\p}^{2} \right) - \frac{\Omega^{2}}{\omega_{0}^2} \left( 1 - 2 \frac{\delta \omega}{\omega_{0}} + 2 i\frac{\Omega \mathcal{B}}{\omega_{0}} \left( 1 + \frac{\rho_{\n}}{\rho_{\p}} \right) \right) \right.} \non
& & {} \left. \left< i 4 \omega_{0} \left( 1 + \frac{\delta \omega}{\omega_{0}} \right) + 8 \mathcal{B} \Omega - \frac{(k_j \Omega^j)^{2}}{\omega_{0}^2 \Omega^{2}} \left( 1 - 2 \frac{\delta \omega}{\omega_{0}} \right)\left[ i 4 \omega_{0} \left( 1 + \frac{\delta \omega}{\omega_{0}} \right) c_{\p}^{2} + 8 \mathcal{B} \Omega c_{\p}^{2} \right] \right> \right\} =0 \ .
\eear
Linearising  (\ref{eq:RotFriN}) we find that
 the mutual friction correction to the waves associated with the neutron fluid is given by 
\bear
\delta \omega
=
2 i \mathcal{B} \Omega \frac{\left[
  \omega_{0}^2
+ 4 \Omega^{2}
\right]
\left[
  \omega_{0}^2
- k^2 \cos^2{\theta} c_{\n}^{2}
\right]}
{
  \omega_{0}^4
+ k^2 \omega_{0}^2 c_{\n}^{2}
+ 4 \Omega^{2} \omega_{0}^2
- 12 k^2 \Omega^{2} \cos^2{\theta} c_{\n}^{2}
} \ .
\eear
Let us first consider the inertial modes, i.e. take $\omega_{0}^2 = 4 \Omega^2 \cos^2{\theta}$. To first order in $\Omega$ this leads to
\bear
\delta \omega
= i \mathcal{B} \Omega \left[
  1
+ \cos^2{\theta}
\right] \ .
\eear
The correction to the sound waves follows by taking $\omega_{0}^2 = k^2 c_{\n}^{2} + 4 \Omega^{2} \sin^{2}{\theta}$.
This leads to
\bear
\delta \omega
= i \mathcal{B} \Omega \sin^2{\theta} \ .
\eear

To find the other set of solutions we linearise (\ref{eq:RotFriP}) and solve for $\delta\omega$. This leads to
\bear
\delta \omega
=
2 i \mathcal{B} \Omega \frac{\rho_{\n}}{\rho_{\p}} \frac{\left[
  \omega_{0}^2
+ 4 \Omega^{2}
\right]
\left[
  \omega_{0}^2
- k^2 \cos^2{\theta} c_{\p}^{2}
\right]}
{
  \omega_{0}^4
+ k^2 \omega_{0}^2 c_{\p}^{2}
+ 4 \Omega^{2} \omega_{0}^2
- 12 k^2 \Omega^{2} \cos^2{\theta} c_{\p}^{2}
} \ .
\eear
For the inertial waves we again use $\omega_{0}^{2} = 4 \Omega^{2} \cos^{2}{\theta}$
and find that,
\be
\delta \omega = i \mathcal{B} \Omega \frac{\rho_{\n}}{\rho_{\p}} \left[ 1 + \cos^{2}{\theta} \right] \ .
\ee
Finally, using the proton sound wave solution $\omega_{0}^{2} = k^{2} c_{\p}^{2} + 4 \Omega^{2} \sin^{2}{\theta}
$ we have
\be
\delta \omega = i \mathcal{B} \Omega \frac{\rho_{\n}}{\rho_{\p}} \sin^{2}{\theta} \ .
\ee

To summarize the results, we now have two sets of sound waves that are damped by mutual friction. Their
frequencies follow from
\bear
\omega & = &  \pm \left( k c_{\n} + 2 \Omega^{2} \sin^{2}{\theta} \right) + i \mathcal{B}\Omega \sin^{2}{\theta} \ , \label{eq:MutualDisipSoundN}\\
\omega & = &  \pm \left( k c_{\p} + 2 \Omega^{2} \sin^{2}{\theta} \right) + i \mathcal{B}\Omega \frac{\rho_{\n}}{\rho_{\p}} \sin^{2}{\theta} \ . \label{eq:MutualDisipSoundP}
\eear
There are also two sets of inertial modes. In the undamped case their frequencies are degenerate, but they become distinct when we account
for the mutual friction. These solutions are
\bear
\omega & = &  \pm 2 \Omega \cos{\theta} + i \mathcal{B} \Omega \left[ 1 + \cos^{2}{\theta} \right] \ , \label{eq:MutualDisipInertN}\\
\omega & = &  \pm 2 \Omega \cos{\theta} + i \mathcal{B} \Omega \frac{\rho_{\n}}{\rho_{\p}} \left[ 1 + \cos^{2}{\theta} \right]
\ . \label{eq:MutualDisipInertP}
\eear
From these results we learn the following.
First of all, (\ref{eq:MutualDisipSoundN}) and (\ref{eq:MutualDisipSoundP}) show that there will be no dissipation of sound waves that travel along the axis of rotation.
This is natural since the sound waves are longitudinal and the mutual friction only affects motion orthogonal to the
vortex array.
It is interesting to contrast this with the result for the inertial modes. From (\ref{eq:MutualDisipInertN}) and (\ref{eq:MutualDisipInertP})
we see that these waves are always damped.
In fact, the effect of the mutual friction is maximal when the wave travels along the vortex array. This result is easy to understand
from the discussion of the single-fluid problem in Section~2.2.  Since the inertial waves generally 
have a component that is orthogonal to the vortex array it is natural
that they experience damping due to mutual friction.

\section{Including vortex tension}

Up to this point we have implicitly assumed that the vortices can be considered straight.
In effect, we have ignored the tension that arises because of  vortex
curvature. This tends to be a relatively small effect, so one would not expect our results
to change much if we account for it. However, it turns out that the vortex tension is
important for the instability that we will discuss in the next section. In particular,
it determines the critical wavelength at which the instability sets in. Hence, it is useful to
extend our  discussion  in such a way that the tension of the neutron vortex array is accounted for.
This discussion is modelled on  Hall's analysis of the corresponding problem in superfluid Helium
\citep{HALLtension}. By redoing his calculation within our formulation we will show
how entrainment affects these modes. For simplicity, we will ignore the mutual friction in this section.
The derivation of the tension term is provided in Appendix~A. Including the relevant contribution,
the equations of motion in a rotating frame
are (as before, we ignore the gravitational potential)
\bear
\left( \pd{}{t} + v^{j}_{\n} \nabla_{j} \right) \left( v_{i}^{\n} + \varepsilon_{\n} w_{i}^{\p\n} \right) + \varepsilon_{\n} w_{j}^{\p\n}
 \nabla_{i} v_{\n}^{j} + \nabla_{i} \tilde{\mu}_{\n} + 2\epsilon_{ijk} \Omega^{j} v^{k}_{\n} & = & \bar{\nu} n_v \kappa^{j} \nabla_{j} \hat{\kappa}_{i} \ ,  \label{mom_ten} \\
\left( \pd{}{t} + v^{j}_{\p} \nabla_{j} \right) \left( v_{i}^{\p} + \varepsilon_{\p} w_{i}^{\n\p} \right) +
\varepsilon_{\p} w_{j}^{\n\p} \nabla_{i} v_{\p}^{j} + \nabla_{i} \tilde{\mu}_{\p} + 2\epsilon_{ijk} \Omega^{j} v^{k}_{\p} & = & 0 \ .
\eear
As discussed in Appendix~A, we have [recall (\ref{momdef})] 
\be
n_v \kappa_i = { 1 \over m_\n } \epsilon_{ijk} \nabla^j p_\n^k = \epsilon_{ijk} \nabla^j  [ v_\n^k + \varepsilon_\n w_{\p\n}^k ] \ , 
\label{nvdef}\ee
while
\be
\bar{\nu} = { 1 - \varepsilon_\p \over 1 - \varepsilon_\n-\varepsilon_\p} \nu  = { 1 - \varepsilon_\p \over 1 - \varepsilon_\n-\varepsilon_\p} { \kappa \over 4 \pi} \log \left( { b \over a_0}\right) \ ,
\ee
see \citet{ASC2} for a detailed discussion.

We have already worked out most of the terms we need to discuss the perturbations of these
equations. The only new piece is the tension contribution. If we consider the same background configuration as in the previous
sections, then the two fluids rotate uniformly at the same rate and we have $n_v\kappa^i = 2\Omega^i$ in the background.
We then need to work out
\be
\delta f_i^\mathrm{tension} = \delta[ \bar{\nu} n_v \kappa^{j} \nabla_{j} \hat{\kappa}_{i}] =
\bar{\nu} \left[ \delta (n_v \kappa^j) \nabla_j \hat{\kappa}_i + 2\Omega^j \nabla_j \delta \hat{\kappa}_i \right] \ .
\ee
The first term is easily worked out from (\ref{nvdef}). The definition also leads to
\be
\delta \hat{\kappa}_i = { 1 \over 2 m_\n \Omega} \left( \epsilon_{ijk} \nabla^j \delta p_\n^k
- \hat{\Omega}_i \hat{\Omega}^l \epsilon_{lmn}\nabla^m \delta p_\n^n \right) \ .
\ee
If the background configuration is uniformly rotating,  we find that
\be
\delta f_i^\mathrm{tension} = { \bar{\nu} \over m_\n} \hat{\Omega}^j \nabla_j \left( \epsilon_{ilm} \nabla^l \delta p_\n^m
- \hat{\Omega}_i  \hat{\Omega}^l \epsilon_{lmn}\nabla^m \delta p_\n^n \right) \ .
\ee

As in the previous sections we now make the plane wave Ansatz, i.e. we assume that $p_\n^i = \bar{p}_\n^i \exp[i(\omega t + k_j x^j)]$.
Then
\be
\delta \bar{f}_i^\mathrm{tension} = - { \bar{\nu} k_z \over m_\n} \left[ \epsilon_{ijk} k^j \bar{p}_\n^k - (\epsilon_{lmn} \hat{\Omega}^l k^m \bar{p}_\n^n) \hat{\Omega}_i \right] \ , 
\label{pert_ten}\ee
where we have defined $k_z = k_j \hat{\Omega}^j$.

Since our main interest is to see how the vortex tension affects the various modes that we have discussed previously,
it is useful to make a further simplification at this point. We will concentrate on waves that propagate along the
axis of rotation. Then $k_z=|k|$ and since $k^i$ is parallel to $\hat{\Omega}^i$ the last term in (\ref{pert_ten})
vanishes.  Hence, we have
\be
\delta \bar{f}_i^\mathrm{tension} = - { \bar{\nu} k \over m_\n}  \epsilon_{ijk} k^j \bar{p}_\n^k = \bar{\nu} k_z^2  \epsilon_{ijk} \hat{k}^j [
\bar{v}_\n^k + \varepsilon_\n (\bar{v}_\p-\bar{v}_\n)] \ .
\ee
This expression  shows that the tension has no effect on longitudinal waves that travel along the
rotation axis. In other words, the sound waves are unaffected by the inclusion of the tension. The same is not true for the inertial waves.

Combining the above results with results from the previous sections we arrive at the perturbed equations of motion;
\be
i \omega [ \bar{v}_{i}^{\n} + \varepsilon_{\n} (\bar{v}^{\p}_{i} - \bar{v}^{\n}_{i})] - i k_{i} \tilde{\mu}_{\n\n} \bar{\rho}_{\n} + 2 \epsilon_{ijk} \Omega^{j} \bar{v}_{\n}^{k} = - \bar{\nu} k_z^2 \epsilon_{ijk} \hat{k}^{j} [\bar{v}^{k}_{\n} + \varepsilon_{\n} (\bar{v}^{k}_{\p} - \bar{v}^{k}_{\n})] \ , 
\ee
and
\be
i \omega [ \bar{v}_{i}^{\p} + \varepsilon_{\p} (\bar{v}^{\n}_{i} - \bar{v}^{\p}_{i})] - i k_{i} \tilde{\mu}_{\p\p} \bar{\rho}_{\p} + 2 \epsilon_{ijk} \Omega^{j} \bar{v}_{\p}^{k} = 0 \ .
\ee

While we could work out the dispersion relation for generic waves in this system, we have chosen 
not to do this. The reason is very simple. As already mentioned, when the wave vector is aligned with the 
rotation axis, as in the above equations, then the 
sound waves are unaffected by the tension. Given this, it is natural to simplify the 
analysis by focussing on pure transverse inertial waves. For transverse waves we have 
$\bar{v}^\X_j k^j=0$ which leads to $\bar{\rho}_\X=0$ by virtue of the continuity equations.
Hence the perturbation equations can be written 
\be
i \omega \bar{p}_{i}^{\n} 
= - 2 \Omega \epsilon_{ijk} \hat{k}^{j} \bar{v}_{\n}^{k} 
 - \bar{\nu} k_z^2 \epsilon_{ijk} \hat{k}^{j} \bar{p}^{k}_{\n} \ , 
\label{eqone}\ee
and
\be
i \omega \bar{p}_{i}^{\p} 
= - 2 \Omega \epsilon_{ijk} \hat{k}^{j} \bar{v}_{\p}^{k} \ .
\label{eqtwo}\ee
To derive the dispersion relation we first take the cross product of each equation with 
$\hat{k}_i$. This leads to the relations
\be
i \omega \varepsilon_{ijk} \hat{k}^j  \bar{p}^{k}_{\n} 
= 2 \Omega \bar{v}^\n_i + \bar{\nu} k_z^2 \bar{p}^\n_i  \ , 
\label{eqthree}\ee
and
\be
i \omega \varepsilon_{ijk} \hat{k}^j  \bar{p}^{k}_{\p}
= 2 \Omega \bar{v}^\n_i \ .
\label{eqfour}\ee
Recalling the definition of the momenta $p_\X^i$ we can solve equations (\ref{eqone}) and 
(\ref{eqtwo}) for $\varepsilon_{ijk}\hat{k}^j \bar{v}_\X^k$. Inserting the results in 
equations (\ref{eqthree}) and (\ref{eqfour}) we have
\begin{eqnarray}
\left\{ [ 2\Omega+(1-\varepsilon_\n)\bar{\nu}k_z^2 ]^2 - \omega^2 (1 - \varepsilon_\n) \left[ 
1 -\varepsilon_\n - {\varepsilon_\n \varepsilon_\p \omega^2 \bar{\nu} k_z^2 \over 2 \Omega} \right] 
- {\varepsilon_\n \varepsilon_\p \omega^2 \over 2 \Omega}
\right\} \bar{v}_i^\n \nonumber \\
= \left\{ \omega^2 \varepsilon_\n (1 - \varepsilon_\n) \left[ 1 - { (1-\varepsilon_\p) \bar{\nu} k_z^2 \over 2 \Omega}
\right] + { \varepsilon_\n (1-\varepsilon_\p)\omega^2 \over 2 \Omega} - [ 2 \Omega + (1 -\varepsilon_\n) \bar{\nu} 
k_z^2] \varepsilon_\n \bar{\nu} k_z^2 \right\} \bar{v}_i^\p \ ,
\end{eqnarray}
and
\be
[4\Omega^2 - (1-\varepsilon_\p)^2 \omega^2 ] \bar{v}_i^\p = \omega^2 \varepsilon_\p^2 \bar{v}_i^\n \ .
\ee
From these two relations we see that the required dispersion relation is
\begin{eqnarray}
[4\Omega^2 - (1-\varepsilon_\p)^2 \omega^2 ]
\left\{ [ 2\Omega+(1-\varepsilon_\n)\bar{\nu}k_z^2 ]^2 - \omega^2 (1 - \varepsilon_\n) \left[
1 -\varepsilon_\n - {\varepsilon_\n \varepsilon_\p \omega^2 \bar{\nu} k_z^2 \over 2 \Omega} \right]
- {\varepsilon_\n \varepsilon_\p \omega^2 \over 2 \Omega}
\right\} \nonumber \\
=
\omega^2 \varepsilon_\p^2
\left\{ \omega^2 \varepsilon_\n (1 - \varepsilon_\n) \left[ 1 - { (1-\varepsilon_\p) \bar{\nu} k_z^2 \over 2 \Omega}
\right] + { \varepsilon_\n (1-\varepsilon_\p)\omega^2 \over 2 \Omega} - [ 2 \Omega + (1 -\varepsilon_\n) \bar{\nu}
k_z^2] \varepsilon_\n \bar{\nu} k_z^2 \right\} \ .
\label{detcon}\end{eqnarray}
In principle, it is straightforward to write down the solutions to this equation. After all, it is just a quadratic in 
$\omega^2$. Of course, the resultant expressions will be so complicated that we learn very little
from them. Let us instead focus on two limiting cases. First of all, we see that if we neglect the 
entrainment we have
\be
[4\Omega^2 - \omega^2 ][  (2\Omega+\bar{\nu}k_z^2 )^2 - \omega^2] =0 \ .
\ee 
The solutions are obviously
\be
\omega = \pm  (2\Omega+\bar{\nu}k_z^2 ) \ ,
\ee
and
\be
\omega = \pm 2 \Omega \ .
\ee
The first solution represents the neutron inertial modes, and the second corresponds to the 
inertial waves in the
proton fluid. As one might expect, the former are affected by the neutron vortex tension 
while the latter are not. These modes  
are analogous to the modes found by
 \citet{HALLtension} in the case of superfluid Helium. Of course, our 
 calculation adds to the standard analysis for Helium by 
 accounting for the entrainment. To get a first idea of how it affects the
 inertial waves, we can include it as a small correction to the above solutions.
 We then find that, to linear order in the entrainment we have
\be
\omega = \pm  (2\Omega+\bar{\nu}k_z^2 + 2 \varepsilon_\n \Omega) \ ,
\ee
and
\be
\omega = \pm 2 (1+\varepsilon_\p) \Omega \ .
\ee
It should, of course, be emphasized here that there is no physical reason why the 
entrainment parameters should be small. We have simply made this assumption in order
to facilitate an analytical calculation.

To summarize, we have shown how the tension of the neutron vortex array provides a
small correction to the inertial modes in the neutron fluid. We have demonstrated that this
remains true when the entrainment is considered weak and the calculation is carried out to linear order.
The full solution to the problem, obtained from (\ref{detcon}), is likely to exhibit a more complex
structure. This could easily be investigated via numerical solutions of
(\ref{detcon}) for some suitable model equation of state. 
At this point we are, however, not going to discuss this possibility. Instead,
we will consider the effect of introducing a relative flow on the background configuration.

\section{Instability of the vortex array}

So far we have assumed that the two fluids rotate together in the background configuration. 
This is a natural assumption given that dissipation will tend to damp any relative motion.
However, there are a number of situations where one may be interested in dynamics that takes place 
on a timescale shorter than that associated with dissipation. Then one can relax the conditions of 
both chemical and dynamical equilibrium. In particular, one can allow for relative 
motion in the background configuration used in the plane-wave analysis. The question is 
whether a relative background flow alters the solution we have discussed in an 
interesting way. This turn out to be the case. In fact, by allowing a relative flow we will 
see that the vortex array may suffer a dynamical instability. This instability is 
well-known in the case of Helium, and is often refered to as the Donnelly-Glaberson instability \citep{Glaberson}.
Since the two-fluid model for a superfluid neutron star core is 
completely analogous to the standard formulation for superfluid Helium, it is no surprise that this
instability is relevant also for neutron stars. In this section we derive the critical relative velocity 
for this vortex instability, and discuss its interpretation. 

In order to keep the analysis tractable we extend the case discussed in the previous section. 
That is, we focus on purely transverse waves in the case when the wave vector $k^i$ is aligned with the rotation $\Omega^i$. 
In addition, we will assume that it is sufficient to consider the dynamics of one of the fluids. 
In practice, we consider the protons as ``clamped'' and ignore their contribution entirely. 
This setup is analogous to that discussed by \cite{Glaberson} for Helium. In that case, the 
assumption can to some extent be justified since the ``normal'' fluid is viscous. In our case, this would also be true, since 
our ``proton'' fluid  accounts for the electron component, which will be affected by viscosity. 
It is not clear, however, that the viscous timescale is short enough that the clamping assumption is truly justified. 
This is an important caveat, but we do not believe that relaxing this assumption would alter our results
in a significant way. For simplicity, we  have also chosen to neglect the entrainment effect here.  

We focus on the perturbed neutron equation in the case when there is a relative flow in the background.
To facilitate the analysis we first assume that this background flow is aligned with both the
wave vector and the rotation axis. Representing the background flow by $V_\n \hat{k}^i$, the 
perturbation equation can be written 
\be
i (\omega + V_\n^j k_j) \bar{v}^\n_i + 2 \Omega \epsilon_{ijk}\hat{k}^j \bar{v}_\n^k = \delta \bar{f}_i \ .
\ee 
The force on the right-hand
side has three contributions. The contribution from the vortex tension remains unchanged from the previous section. 
We also need the mutual friction force. Under the present assumptions, and if we also neglect $\mathcal{B}'$ which 
makes sense since it is much smaller than $\mathcal{B}$,  we get from (\ref{fmf}),
\be
\delta \bar{f}_i^\n = \mathcal{B} \epsilon_{ijk} \epsilon^{klm} \left[ 
i \hom_l V_m^\n \epsilon^{jpq} k_p \bv^\n_q 
 + i \hom^j \epsilon_{lpq} k^p \bv_\n^q V_m^\n + 2 \Omega^j \hom_l \bv_m^\n  \right] \ .
\ee 
Finally, we also want to  account for the contribution to the mutual friction from the self-induced flow. 
This is a small contribution, but it is natural to included it if we are considering the vortex tension. 
As long as the background flow is uniform, this term can be written, cf. Eq.~(36) in \cite{ASC2},
\be
f_i^\mathrm{ind} = - \nu \mathcal{B} n_v  \epsilon_{ijk} \kappa^j \hat{\kappa}^l \nabla_l \hat{\kappa}^k \ . 
\label{induced}\ee
Perturbing this we arrive at the contribution 
\be
\delta \bar{f}_i^\mathrm{ind} =  \nu \mathcal{B} (\hom^l k_l) \epsilon_{ijk} \hom^j
\epsilon^{kmn} k_m \bv^\n_n \ .
\ee

Putting all this together, we consider an equation of form
\be
i (\omega + V_\n k_z) \bar{v}^\n_i + 2 \Omega \epsilon_{ijk}\hat{k}^j \bar{v}_\n^k = 
- (\nu k_z^2 - i\mathcal{B} V_\n k_z) \epsilon_{ijk} \hat{k}^j \bar{v}_\n^k - \mathcal{B} (2 \Omega + \nu k_z^2 ) \bar{v}^\n_i
\ . 
\label{ppwave}\ee
Taking the cross product of this equation with $\hat{k}^i$ and combining the two equations we find that the
dispersion relation is simply
\be
[ \omega + V_\n k_z - i \mathcal{B} (2 \Omega + \nu k_z^2)]^2 =
[2 \Omega + \nu k_z^2 \mp V_\n k_z ]^2 \ .
\ee
That is, the inertial waves in this system must have frequency
\be
\omega + V_\n k_z = \pm (2 \Omega + \nu k_z^2) + i \mathcal{B} (2 \Omega + \nu k_z^2 \mp V_\n k_z)  \ .
\ee
Given that our assumed time-dependence is $\exp(i\omega t)$ this expression shows that the solution 
corresponding to the upper sign will be exponentially growing ($\omega$ has a negative imaginary part) 
when 
\be
V_\n > { 2 \Omega \over k_z} + \nu k_z \ .
\label{vcrit}\ee
In other words, for any given wavevector $k_z$ there exists a critical relative flow above which 
the wave is unstable. Of course, we see from (\ref{vcrit}) that the critical flow must 
be large both in the limits of large and small $k_z$. If we are interested in the critical 
flow at which the instability first sets in, then we simply need to find the minimum of
the function on the right-hand side of (\ref{vcrit}). Thus we need 
\be
- { 2 \Omega \over k_z^2} + \nu = 0 \quad \longrightarrow \quad k_z = \sqrt{ {2 \Omega \over \nu} } \ . 
\ee 
Inserting this in the expression for the critical flow we see that the system will have 
unstable waves when
\be
V_\n > V_c = 2 \sqrt{ 2 \Omega \nu} \ .
\ee
This is exactly the condition derived by \citet{Glaberson} for the Helium problem.
 
Even though we have not attempted the general problem, without assuming that the protons are clamped, 
we have relaxed some of the other assumptions. In particular, one does not have to assume that the waves are purely
transverse. The interested reader can find a more general discussion in Appendix~B.

Let us now see if we can understand the nature of this instability better. To do this it is helpful to 
consider the phase-velocity of the waves. Recall that in the present problem setup, 
a constant phase would mean that
\be
\mathrm{Re}\ \omega t + k_z z = \mathrm{constant} \quad
\longrightarrow \sigma_p = - {\mathrm{Re}\ \omega \over k_z} \ , 
\ee
where $\sigma_p$ is the phase-velocity. Hence the phase-velocity of the inertial waves
is
\be
\sigma_p = V_\n \mp  (2 \Omega + \nu k_z^2) \ .
\ee
Comparing this to the condition for the critical velocity we immediately see that the 
instability sets in through the waves that propagate in the direction opposite the background flow,
(for a suitably small $V_\n$).
The critical point is simply identified as $\sigma_p=0$. The interpretation of this condition is that a 
wave that is originally seen as travelling downwards (relative to $V_\n$) is dragged upwards 
by the flow and becomes unstable when its direction of propagation changes (according to a fixed observer). 
This condition is typical for a  two-stream instability. We have previously considered this class of 
instabilities for neutron stars, see \citet{twostream} for a discussion and a list of 
relevant references to the 
plasma physics literature. A two-stream instability typically requires two identifiable flows and some coupling 
between them. In our previous discussion of such instabilities for neutron stars, we focussed on 
chemical coupling and the role of entrainment. We now see that the instability can also be caused 
by the mutual friction. This possibility is particularly interesting since the instability may be intimately linked to 
the formation of vortex loops and superfluid turbulence \citep{ASC2}. In fact, the present analysis provides
an important complement to our previous discussion of the turbulence problem.

It is obviously necessary to ask whether this instability is likely to operate in neutron stars. 
For this to be the case, one would require the critical wavelength to be much smaller that (say) the size of the star. 
Otherwise, the plane-wave analysis does not apply. From our previous discussion \citep{ASC2}  we know that 
\be
\nu = { \kappa \over 4 \pi} \log \left( { b \over a_0} \right) \ , 
\ee 
where $\kappa \approx 2\times 10^{-3}$~cm$^2$/s and $\log (b/a_0) \approx 20$. 
From this we see that the critical wavelength for which the instability first appears is
\be
k_z \approx 250 \left( {\Omega \over 100\ \mbox{s}^{-1}} \right)^{1/2} \ \mbox{cm}^{-1} \ ,
\ee 
corresponding to a wavelength
\be
\lambda =  { 1 \over k_z} \approx 4 \times 10^{-3}  \left( {\Omega \over 100\ \mbox{s}^{-1}} \right)^{-1/2} \ \mbox{cm} \ .
\ee
If we compare this to the typical intervortex spacing
\be
b \approx 3.4 \times 10^{-3} \left( {\Omega \over 100\ \mbox{s}^{-1}} \right)^{-1/2} \ \mbox{cm} \ ,
\ee
we conclude that one may well expect modes with a wide range of wavelengths to be unstable in a typical neutron star.  
This is an interesting possibility, and it would be exciting to consider various scenarios 
where the instability may operate. 

\section{Turbulent mutual friction}

The presence of a dynamical instability in the vortex array will lead to oscillations 
in the vortices, triggering reconnections and the formation of vortex loops 
with a range of different sizes \citep{ASC2}. This behaviour is very similar to the standard 
cascade seen in normal fluid turbulence. If a turbulent tangle is present, then 
our analysis is no longer valid. After all, the form we are using for the mutual 
friction force is based on the assumption that the vortex array is (locally) straight. 
One of the outstanding issues in superfluid Helium research concerns the nature
of the force in the turbulent case. While some sort of consensus has been reached in the case of isotropic 
turbulence, problems with both relative flow and rotation are still far from understood.
Yet this is the problem that we need to solve in order to model neutron stars. Our system 
is rotating, and if it becomes turbulent then any tangle that develops should be polarised. 

In absence of a clear strategy for developing a detailed model for the mutual friction 
force in the case of polarised turbulence, we have previously proposed a phenomenological 
prescription \citep{ASC2}.  It is interesting to apply this decription to the plane-wave problem we
are currently investigating, since this may lead to a better understanding of the effect that turbulence
may have on the vortex instability. 

We take as our starting point the mutual friction force posited in Eq.~(78) of \citet{ASC2}.  
In essence, this means that we add a term accounting for the presence of a turbulent tangle to 
(\ref{fmf}) which now represents to polarisation of the vortex array. 
Thus we have 
\be
f_i^\mathrm{mf}  = \ldots + { 2 L_T \over 3} \kappa \mathcal{B} w_i^{\p\n} \ ,
\ee
where the $\ldots$ represents the straight vortex term from before.
We have defined the total vortex length per unit volume, 
\be
L_T =  \left( {\alpha_1 \over \beta} { W \over \kappa}\right)^2
+ { \alpha_1 \beta_1 \over \beta^2} {W \over \kappa} L_R^{1/2} + \left[ { 1 \over 4} \left( {\beta_1 \over \beta}\right)^2 -1 \right]L_R \ ,
\ee
where $W= | w_{\n\p}^i \hat{\kappa}_i| = V_\n$, $L_R=n_v$ and the constant parameters are such that (the arguments for this are
given by \citet{ASC2})
\be
{ \alpha_1 \over \beta} \approx \chi { \mathcal{B} \over 2 \pi} \ , \qquad \mbox{ and } \quad { \beta_1 \over \beta} \approx 2 \ ,
\label{params}\ee
with $\chi$ of order unity.
If we work out the perturbation of this new contribution to the overall force in the particular case when the 
wave vector is aligned with the background rotation, then we find that $\delta L_R=0$. Moreover, when the imposed relative flow is
also aligned with  the rotation, we have $\delta W=0$. This makes it very easy to account for this new force contribution. 
Under the conditions assumed in the previous section (protons clamped etcetera), we simply get
\be
\delta \bar{f}_i^\mathrm{mf} = \ldots - { 2 L_T \over 3} \kappa \mathcal{B} \bar{v}_i^n \ .
\ee
When this term is added to the right-hand side of (\ref{ppwave}), and the dispersion relation is 
worked out as before, we find the wave solutions
\be
\tilde{\omega} = \pm (2 \Omega + \nu k_z^2) + i \mathcal{B} (2 \Omega + \nu k_z^2 \mp V_\n k_z) 
+ { 2 i L_T \over 3} \kappa \mathcal{B} \ .
\label{mode_final}\ee
This  suggests that turbulence always damps the inertial waves, as one might have expected. 
For the parameters given in (\ref{params}) we see that
\be
L_T = \left( { \chi \mathcal{B} V_\n \over 2 \pi \kappa } \right)^2 + \chi {\mathcal{B} V_\n \over \pi \kappa } \left( {2 \Omega \over \kappa} \right)^{1/2}  \ .
\ee
This shows that the new damping term in (\ref{mode_final}) is very small in the neutron star case, when 
$\mathcal{B} \ll 1$. If we nevertheless include this contribution, and work out the critical velocities we 
 find two roots. Assuming that $k_z$ is suitably large we retain (\ref{vcrit}) as the velocity at which the 
instability sets in. In addition, we find a second critical flow, beyond which the system is stable. The 
corresponding critical velocity is approximately given by
\be
V_\n \approx { 6 \pi^2 \over \chi \mathcal{B}} \kappa k_z \ .
\ee  
For typical parameters, this velocity would be vastly greater than the critical velocity at which the 
instability sets in. In fact, it may well be the case that one can not reach such large relative flows in a realistic neutron star.  
Nevertheless, the result is conceptually interesting. One should also keep in mind that $\mathcal{B}$ is of order unity in superfluid
Helium \citep{donnelly}, so this upper cut-off for the vortex instability may not be out of reach in that context.

\section{Brief summary}

We have analysed the wave propagation in a rotating superfluid neutron star cores, 
taking into account the standard mutual friction force.  
Our plane-wave analysis has added to previous discussions of this problem in a number of important ways. 
First of all, for models where the two background fluids co-rotate,  we have clarified the role 
of chemical coupling and entrainment on both sound and inertial waves. Secondly, we have considered the
 mutual friction damping, demonstrating the well-known fact that 
sound waves propagating along a vortex array are undamped. We have also shown that the same is not true 
for inertial waves, which are damped by the mutual friction regardless of the propagation direction.
We have accounted for the relatively small contribution of the  vortex tension, which arises due to local vortex curvature.
Focussing on purely transverse inertial waves, we derived the correction that the tension induces in the wave 
frequency. 

The most exciting result of our investigation concerns the presence of a dynamical instability associated with the inertial waves. 
The instability requires a linear relative  flow in the background. We analysed the particular 
case when this flow is aligned with the rotation axis. 
This led to a demonstration that the mutual friction coupling induces an instability once the relative
velocity has reached a critical level. 
This instability is well-known from the analogous problem for superfluid Helium, and  hence our result 
should not come as a great surprise. Nevertheless, the possibility that this instability may operate in neutron stars 
has only recently been appreciated \citep{peralta1,peralta2,ASC2}. We have argued that the instability belongs to the general 
class of two-stream instabilities. This interpretation is (we believe) new, and adds insight also into the 
Helium problem. 

If this instability operates in a neutron star, it is likely to lead to the formation of a 
vortex tangle and a state of superfluid turbulence. The impact of this on, for example, glitch recovery 
is not yet understood. Nevertheless,  it is clear that much of our current ``understanding'' (which tends to be based on the assumption 
of a locally straight vortex array) may have to be revised. 
In view of this, the 
results we have presented here are exciting. Having said that, it is clear that there are a number of difficult issues that 
need to be addressed if we really want to understand this problem. Our analysis was based on a number of simplifying assumptions,
in particular we assumed that the proton fluid was clamped. It would be relevant to try to consider the general problem. 
One would certainly want to account for the entrainment, which will alter the critical velocity for the onset of the instability etcetera.
It would  also be relevant to try to quantify the damping (and possibly stabilising role) of shear viscosity, which should be important 
for short wavelength oscillations.  
We also need to consider various astrophysical scenarios for which the instability may be relevant. 
If it is the case that the key features required are a straight vortex array and some imposed relative flow, then the 
instability could be relevant in a number of situations. The most obvious possibilities would be i)  
neutron star free precession where the neutrons and protons essentially rotate with respect to different axes (in the 
simplest model),  ii)  neutron star spin-down which (in a non-magnetic star) is faciliated via a 
viscous Ekman layer at the base of the crust inducing a global flow in the charged component, and iii)
  global mode oscillations, where the length scale of the mode is vastly larger than the typical lengthscale of the 
instability. These are all exciting problems, well worthy of further consideration.

\section*{Acknowledgments}

This work was supported by PPARC through grant numbers PPA/G/S/2002/00038 and
PP/E001025/1.
NA also acknowledges support from PPARC via Senior Research Fellowship no
PP/C505791/1. GLC acknowledges partial support from NSF via grant number 
PHY-0457072.

\appendix 
\section{The vortex tension}

In this Appendix we provide the argument that leads to the form for the neutron vortex tension used in the
main body of the paper.
The calculation is based on the intuitive reasoning of, in particular, \citet{HALLtension}.
It is important in the sense that it demonstrates how the entrainment
parameters enter in the vortex tension.

The starting point is the conservation of vorticity. Defining the macroscopic
vorticity as
\be
\omega^i = n_v \kappa^i = { 1 \over m_\n} \epsilon^{ijk} \nabla_j p^\n_k \ ,
\ee
where the neutron momentum is $p_\n^i = m_\n \left[ v_\n^i + \varepsilon_\n(v_\p^i - v_\n^i) \right]$, 
we have
\be
{ D \over Dt} \int_V n_v \kappa_i dV + \int_S n_v \kappa_i v_L^j dS_j = 0 \ .
\ee
Here it is assumed that the vortices move collectively with velocity
$v_L^i$.
Use the divergence theorem to see that we must have
\be
\partial_t \omega_i + \nabla_j(\omega_i v_L^j) = 0 \ .
\ee
Now note that
\begin{eqnarray}
\nabla_j \omega^j = 0 \ , \\
\omega^j \nabla_j v_L^i = 0 \ .
\end{eqnarray}
The first statement is trivial given the definition of the vorticity. The second
should be true provided that there is no motion along the vortices themselves.
This way the above conservation law can be recast as
\be
\partial_t \omega_i + \nabla^j (\epsilon_{ijk} \epsilon^{klm} \omega_l v^L_m) = 0 \ .
\ee
This leads to
\be
\epsilon_{ijk} \nabla^j \left\{ \partial_t p_\n^k - \epsilon^{klm} v_l^L \epsilon_{mno} \nabla^n p_\n^o \right\} = 0 \ ,
\ee
which then requires that
\be
\partial_t p_\n^k - \epsilon^{klm} v_l^L \epsilon_{mno} \nabla^n p_\n^o = \nabla^k \Psi \ ,
\label{peq}\ee
with $\Psi$ some scalar potential.

Let us now, for simplicity, assume that the only force that acts on the vortex is the
Magnus force. Then we must have
\be
v^L_i = v^\n_i + v^\mathrm{ind,n}_i \ .
\ee
The first term represents the smooth irrotational flow past the vortex, due to
for instance the presence of all other vortices. The second term represents
the self-induced flow that arises when the vortex is curved [see Appendix of \citet{ASC2}]. As we have shown
elsewhere, this term can be written
\be
v_i^\mathrm{ind,n} = { 1 -\varepsilon_\p \over 1 - \varepsilon_\n - \varepsilon_\p} \nu \epsilon_{ijk} \hat{\kappa}^j
\hat{\kappa}^l \nabla_l \hat{\kappa}^k = \bar{\nu} \epsilon_{ijk} \hat{\kappa}^j
\hat{\kappa}^l \nabla_l \hat{\kappa}^k \ .
\ee
%Due to entrainment, there will also be a flow in the protons. We have
%\be
%v_i^\mathrm{ind,p} = - {\varepsilon_\p \over 1 - \varepsilon_\n - \varepsilon_\p} \nu \epsilon_{ijk} \hat{\kappa}^j
%\hat{\kappa}^l \nabla_l \hat{\kappa}^k = \tilde{\nu} \epsilon_{ijk} \hat{\kappa}^j
%\hat{\kappa}^l \nabla_l \hat{\kappa}^k
%\ee

In order to use this vortex velocity in the equation of motion (\ref{peq}), we note that
\be
\epsilon^{klm} \left( v^\n_l + v^\mathrm{ind,n}_l \right) \kappa_m = \varepsilon^{klm} v_l^\n \kappa_m + \bar{\nu} \kappa^l \nabla_l \hat{\kappa}^k \ .
\ee
Then we need
\be
m_\n n_v \epsilon^{klm} \left( v^\n_l + v^\mathrm{ind,n}_l \right) \kappa_m
= v^\n_l \nabla^k p_\n^l - v^\n_l \nabla^l p_\n^k + m_\n n_v \bar{\nu} \kappa^l \nabla_l \hat{\kappa}^k \ .
\ee
Use this in (\ref{peq}) to get
\be
\partial_t p_\n^k + v_l^\n \nabla^l p_\n^k - v_l^\n \nabla^k p_\n^l = \nabla^k \Psi +
m_\n n_v \bar{\nu} \kappa^l \nabla_l \hat{\kappa}^k \ .
\ee
Finally use the definition of the momentum to get
\be
(\partial_t + v_\n^j \nabla_j )(v_\n^i + \varepsilon_\n w_{\p\n}^i) +
\varepsilon_\n w_{\p\n}^j \nabla^i v_j^\n = \nabla^i \chi +
 n_v \bar{\nu} \kappa^j \nabla_j \hat{\kappa}^k \ ,
\ee
where we recall that the velocity difference is
$w_{\p\n}^i = v_\p^i - v_\n^i$.
This is the equation of motion for the superfluid neutrons, with the
contribution from the vortex tension accounted for. The scalar potential $\chi$ can easily be
interpreted as the sum of the chemical and gravitational potentials to arrive at the standard form
for this term. This way we arrive at Eq.~(\ref{mom_ten}) in the main text.

\section{The vortex instability in a more general context}

In this Appendix we provide a slightly more general derivation of the 
vortex two-stream instability 
that was dicussed in Section~6. While we still assume that the proton 
fluid is clamped, and  neglect entrainment, we  
initially relax the assumption that the wave vector is aligned with the rotation 
axis. We also do not assume that the waves are purely transverse. 
The results obtained in Section~6 follow in the appropriate limits, 
and the more complicated calculation that we outline here shows how the 
instability threshold can be derived under less constrained conditions.

In the general case, the plane wave equation for the neutron fluid 
can be written, cf. (\ref{fmf}), 
\be
i \tilde{\omega} \bar{v}^\n_i + i k_i \bar{\mu}_\n + 2 \epsilon_{ijk} \Omega^j \bar{v}_\n^k = 
\delta \bar{f}^\n_i \ ,
\ee
where $\tilde{\omega}= \omega + V_n^j k_j$ and we have used $\delta \tilde{\mu}_\n = \bar{\mu}_\n e^{i(\omega t + k_j x^j)}$.
The force $\delta \bar{f}^\n_i$ is made up of three contributions. 
The first is the mutual friction for a straight vortex array, and it leads
to
\be
\delta \bar{f}_i^\n = \mathcal{B} \epsilon_{ijk} \epsilon^{klm} \left[ 
i \hom_l V_m^\n ( \epsilon^{jpq} k_p \bv^\n_q - \hom^j \epsilon_{pqr} \hom^p k^q \bv_\n^r)
 + i \hom^j \epsilon_{lpq} k^p \bv_\n^q V_m^\n + 2 \Omega^j \hom_l \bv_m^\n  \right] \ .
\ee 
Next we have the contribution from the self-induced flow, which accounts for the vortex curvature. 
As long as the background flow is uniform,  we can perturb (\ref{induced}) to get the contribution 
\be
\delta \bar{f}_i^\mathrm{ind} =  \nu \mathcal{B} (\hom^l k_l) \epsilon_{ijk} \hom^j \left[ 
\epsilon^{kmn} k_m \bv_\n^n - \hom^k \epsilon_{pqr} \hom^p k^q \bv_\n^r \right] \ .
\ee
Finally, we have the vortex tension which is given by (\ref{pert_ten}), i.e. 
\be
\delta \bar{f}_i^\mathrm{ten} = - \nu (\hom^j k_j) \left[ \epsilon_{ilm} k^l \bv_\n^m 
- \hom_i (\epsilon_{lpq} \hom^l k^p \bv_\n^q) \right] \ .
\ee

Putting all the pieces together and rearranging, the perturbed momentum equation can be written
\begin{eqnarray}
\left\{ \tilde{\omega} - i \mathcal{B} [2 \Omega + \nu (k_j \hat{\Omega}^j)^2]\right\} \bar{v}^\n_i 
+ \left[  \bar{\mu}_\n + i \nu \mathcal{B} (\hat{\Omega}^j \bar{v}^\n_j)(\hat{\Omega}^l k_l)\right] k_i
- 2 i \epsilon_{ijk} \Omega^j \bar{v}_\n^k \nonumber \\
= 
\left\{ [ \mathcal{B} V^\n_j -\mathcal{B} (V^\n_n\hat{\Omega}^n) \hat{\Omega}_j - i \nu (k_n \hat{\Omega}^n)]
\epsilon^{jkl}k_k \bar{v}^\n_l - 2i\mathcal{B} \Omega (\hat{\Omega}^j \bar{v}^\n_j)
\right\} \hat{\Omega}_i \nonumber \\
- \mathcal{B} (\epsilon_{jlm} \hat{\Omega}^j k^l \bar{v}_\n^m) V^\n_i +
[ \mathcal{B} (V^\n_j \hat{\Omega}^j) + i \nu (k_j \hat{\Omega}^j)] \epsilon_{ilm} k^l \bar{v}_\n^m \ .
\label{big}\end{eqnarray}

Here the chemical potential perturbation is (in the clamped case we are also assuming that the proton
density variation vanishes) given by
\be
 \bar{\mu}_\n = { \partial \tilde{\mu}_\n \over \partial \rho_\n} \bar{\rho}_\n \ .
\ee
Since the continuity equation gives
\be
i \tilde{\omega} \bar{\rho}_\n + i\rho_\n (k_j \bar{v}_\n^j) = 0  \ ,
\ee
we get,  using the standard definition of the sound speed from (\ref{c_sound}), 
\be
 \bar{\mu}_\n = - { c_\n^2 \over \tilde{\omega}} (k_j \bar{v}_\n^j) \ .
\ee

The trick now is to form different scalar equations from (\ref{big}). 
By taking the scalar product with $k^i$ we get
\begin{eqnarray}
\left[ \tilde{\omega} - i \mathcal{B} (2 \Omega + \nu k_z^2) - { c_\n^2 k^2 \over \tilde{\omega}} \right] (k_i \bar{v}_\n^i)
+ i k_z \mathcal{B} (2\Omega + \nu k^2) (\hat{\Omega}^j \bar{v}^\n_j) \nonumber \\
+  \left[ 2 i \Omega + \mathcal{B} k_z (V_n^\n \hat{\Omega}^n) + i\nu k_z^2 - \mathcal{B} (k_n V^n_\n) \right] W =
 \mathcal{B} k_z (\epsilon^{jkl}V^\n_j k_k \bar{v}^\n_l) \ .
\label{two}\end{eqnarray}
In writing down this expression we have decomposed the wave vector into a piece along the rotation axis and a piece 
orthogonal to it, i.e., we are using
\be
k^i = k^i_z + k^i_\perp \ , \qquad \mbox{ where } \quad k^j_\perp \Omega_j = 0 \ .
\ee
We have also defined the scalar quantity $W = \epsilon_{ijk}\hom^i k^j \bv_\n^k$.

If we take the scalar product of (\ref{big}) with $\hat{\Omega}$ we find another scalar relation;
\be
\tilde{\omega} (\hom^i\bv_i^\n) - {k_z c_\n^2 \over \tilde{\omega}} (k_j \bv_\n^j)
+ \mathcal{B} (V_j^\n \hom^j) W
= \mathcal{B} (\epsilon^{jkl} V^\n_j k_k \bv^\n_l) \ .
\label{three}\ee
From the combination (\ref{two})$-k_z\times$(\ref{three}) we then get
\begin{eqnarray}
\left[ \tilde{\omega} - i \mathcal{B}(2\Omega+\nu k_z^2) - { c_\n^2 \over \tilde{\omega}} (k^2 -k_z^2)
 \right] (k_j \bv_\n^j) + 
 \left[ i\mathcal{B} k_z (2\Omega + \nu k^2) - k_z\tilde{\omega} \right] (\hom_i \bv_\n^i) \nonumber \\
 + [ i(2\Omega + \nu k_z^2) + \mathcal{B} (k_j V_\n^j) ] W = 0
\ .
\label{six}\end{eqnarray}

By taking the cross product between $k^i$ and (\ref{big}) we get another useful relation.
After some work it can be written
\begin{eqnarray}
[i k_z (2\Omega + \nu k^2) + \mathcal{B} k^2 (V^\n_j \hom^j)] \bv^\n_i 
- [ \mathcal{B}(V^\n_j \hom^j) +i\nu k_z] (k_l \bv_\n^l) k_i \nonumber \\
= 
2i (k_j \bv_\n^j) \Omega_i 
- [ \tilde{\omega} - i \mathcal{B} (2 \Omega + \nu k_z^2)] \epsilon_{imn}k^m \bv^n \nonumber \\
+ \left\{ [ \mathcal{B} V^\n_j -\mathcal{B} (V^\n_n \hom^n) \hom_j - i\nu k_z \hom_j] \epsilon^{jkl} k_k \bv^\n_l  - 2 i \mathcal{B} \Omega (\hom^j
\bv^\n_j)\right\} \epsilon_{imp} k^m \hom^p \nonumber \\
- \mathcal{B} W \epsilon_{ipq}k^p V_\n^q \ .
\label{big2}\end{eqnarray}

Taking the scalar product of (\ref{big2}) with $\hom^i$  we arrive at 
\begin{eqnarray}
[ik_z(2\Omega +\nu k^2) +\mathcal{B} k^2 (\bv_j^\n \hom^j)] (\hom_i \bv_\n^i)
-[ i(2\Omega+\nu k_z^2) +\mathcal{B} k_z (\hom_j V_\n^j)] (k_l \bv_\n^l) \nonumber \\
+ [ \tilde{\omega} -i \mathcal{B}(2\Omega + \nu k_z^2) + \mathcal{B} \bar{\Omega} ] W = 0 \ .
\label{five}\end{eqnarray}
Here we have defined yet another scalar
$\bar{\Omega} = \epsilon_{ijk}\hom^i k^j V_\n^k$.

We now have three equations, (\ref{two}), (\ref{six}) and (\ref{five}), for four unknown scalar quantities. 
To solve the general problem we  need another relation. Although this relation can be obtained 
in a few steps by following the above strategy, we choose not to write it down here. 
Instead, we note that if we were to 
align $V_\n^i$ with $\hom^i$ then\footnote{We did not assume alignment from the beginning since 
we wanted to outline how the general problem would be solved.}
\begin{eqnarray}
\bar{\Omega} &=& 0 \ , \\
\epsilon_{ijk} V_\n^i k^j \bv_\n^k &=& V_\n W \ .
\end{eqnarray}  
Then (\ref{three}) simplifies to
\be
\tilde{\omega} (\hom_i \bv_\n^i) - { k_z c_\n^2 \over \tilde{\omega}} (k_j \bv_\n^j)  = 0  \ .
\ee
We can use this relation to get an expression for $k_j \bv_\n^j$. Using the result in 
(\ref{six}) and (\ref{five}) we only have two equations to solve. They are
\begin{eqnarray}
\left\{ {\tilde{\omega}^2 \over k_z c_\n^2} \left[ \tilde{\omega} - i\mathcal{B} (2\Omega + \nu k_z^2)
- { c_\n^2 \over \tilde{\omega}} (k^2 - k_z^2)\right] 
+ k_z [ i \mathcal{B} (2\Omega + \nu k^2) - \tilde{\omega} ] \right\} (\hom_i \bv_\n^i) \nonumber \\
+ [i (2\Omega + \nu k_z^2) + \mathcal{B} k_z V_\n] W = 0 \ , 
\end{eqnarray}
and
\begin{eqnarray}
\left\{ - { \tilde{\omega}^2 \over k_z c_\n^2} [i( 2\Omega + \nu k_z^2) + \mathcal{B} k_z V_\n ] +ik_z (2\Omega + \nu k^2) + \mathcal{B} k^2 V_n  \right\} (\hom_i \bv_\n^i) \nonumber \\
+ [ \tilde{\omega} - i \mathcal{B}(2 \Omega + \nu k_z^2)] W =0  \ .
\end{eqnarray}
From this we easily obtain the dispersion relation
\begin{eqnarray}
\left\{ {\tilde{\omega}^2 \over k_z c_\n^2} \left[ \tilde{\omega} - i\mathcal{B} (2\Omega + \nu k_z^2)
- { c_\n^2 \over \tilde{\omega}} (k^2 - k_z^2)\right]
+ k_z [ i \mathcal{B} (2\Omega + \nu k^2) - \tilde{\omega} ] \right\} 
[ \tilde{\omega} - i \mathcal{B}(2 \Omega + \nu k_z^2)] \nonumber \\
- \left\{  { \tilde{\omega}^2 \over k_z c_\n^2} [( 2\Omega + \nu k_z^2) 
- i\mathcal{B} k_z V_\n ] - k_z (2\Omega + \nu k^2) +i \mathcal{B} k^2 V_n  \right\}
[(2\Omega + \nu k_z^2) -i \mathcal{B} k_z V_\n] =0 \ .
\end{eqnarray}

It is, of course, not easy to write down the general solutions to this dispersion relation. But we can learn a lot from it if we 
make some further simplifications. To discuss these examples we first note that $k_z = k \cos \theta$, where 
$\theta$ is the angle between the wave vector and the rotation axis.  It is then straightforward to verify that we retain the 
solution from section~5 in the case when  $\theta= \mathcal{B}=0$. We obviously also get the neutron sound waves. 
If we focus our attention on the possible vortex instability, then it would be natural to first relax the assumption that $\theta$ vanishes. 
Doing this, but still leaving $\mathcal{B}=0$ (and in addition assuming slow rotation and weak tension) we find the leading order
wave solutions;
\begin{eqnarray}
\tilde{\omega}^2= \left\{ \begin{array}{ll}
\pm k c_\n\left[ 1  + (\Omega \sin^2\theta/ k^2 c_n^2) (2\Omega + \nu k^2 \cos^2 \theta)  \right] \ ,  \qquad \mbox{sound waves} \ ,\\
\  \\
\pm \cos \theta (2 \Omega + \nu k^2)^{1/2} ( 2 \Omega + \nu k^2 \cos^2 \theta)^{1/2} \ , \qquad \mbox{inertial waves} \ .
\end{array}\right. 
\end{eqnarray} 
If we now linearise the dispersion relation in $\mathcal{B}$, and assume that the modes take the form 
$\tilde{\omega} = \tilde{\omega}_0 + \mathcal{B} \delta \tilde{\omega}$, then we find that the 
mutual friction induced frequency correction follows from
\be
\delta \tilde{\omega} =  i {\tilde{\omega}_0 -k V_\n \cos \theta \over \tilde{\omega}_0} \times
\left\{ \begin{array}{ll}  \Omega \sin^2 \theta \ , \qquad \mbox{sound} \ , \\ \\
\Omega ( 1 + \cos^2 \theta) + \nu k^2 \cos^2 \theta
 \ , \qquad \mbox{inertial}\ . \end{array} \right.
\ee
Recalling that the waves are unstable if the imaginary part is negative,  
we see that (assuming that $k \cos \theta \ge 0$) the solutions for which $\tilde{\omega}_0<0$ are always stable. 
In contrast, the $\tilde{\omega}_0>0$ solutions become unstable at the critical velocity
\be
V_c = { \tilde{\omega}_0 \over k \cos \theta} \ .
\ee
As one might have guessed, the onset of the instability depends on the projection of the 
wave vector along the relative flow. For the sound waves we thus find that the critical flow is 
\be
V_c^\mathrm{sound} \approx { c_\n \over \cos \theta} \ .
\ee 
Since the superfluidity is likely broken before the wave propagation reaches 
the speed of sound, this indicates that these modes are always stable in a real system. 
Again the inertial waves are different. We find that
\be
V_c^\mathrm{inertial} \approx   { 1 \over k} (2 \Omega + \nu k^2)^{1/2} ( 2 \Omega + \nu k^2 \cos^2 \theta)^{1/2} \ .
\ee
According to this criterion, the instability actually sets in at a lower relative velocity when 
the wave vector is not aligned with the rotation axis. Of course, in reality one may expect the tension term to be small 
compared to the rotation term. Then the difference between the above result and the aligned case discussed in Section~6 
may only be significant
at extremely short wavelengths.

\end{document}